\input harvmac.tex
 \input epsf.tex
 \input amssym
 \input color
 \input

\def\figin{\epsfcheck\figin}\def\figins{\epsfcheck\figins}
\def\epsfcheck{\ifx\epsfbox\UnDeFiNeD
\message{(NO epsf.tex, FIGURES WILL BE IGNORED)}
\gdef\figin##1{\vskip2in}\gdef\figins##1{\hskip.5in}
\else\message{(FIGURES WILL BE INCLUDED)}%
\gdef\figin##1{##1}\gdef\figins##1{##1}\fi}
\def\DefWarn#1{}
\def\figinsert{\goodbreak\midinsert}
\def\ifig#1#2#3{\DefWarn#1\xdef#1{fig.~\the\figno}
\writedef{#1\leftbracket fig.\noexpand~\the\figno} %
\figinsert\figin{\centerline{#3}}\medskip\centerline{\vbox{\baselineskip12pt
\advance\hsize by -1truein\noindent\footnotefont{\bf
Fig.~\the\figno:} #2}}
\bigskip\endinsert\global\advance\figno by1}


\def \pa {\partial}

\def \eps {\epsilon}

\lref\VenezianoYB{
  G.~Veneziano,
  ``Construction of a crossing - symmetric, Regge behaved amplitude for linearly rising trajectories,''
Nuovo Cim.\ A {\bf 57}, 190 (1968).
}

\lref\FitzpatrickYX{
  A.~L.~Fitzpatrick, J.~Kaplan, D.~Poland and D.~Simmons-Duffin,
``The Analytic Bootstrap and AdS Superhorizon Locality,''
JHEP {\bf 1312}, 004 (2013).
[arXiv:1212.3616 [hep-th]].
}

\lref\KomargodskiEK{
  Z.~Komargodski and A.~Zhiboedov,
  ``Convexity and Liberation at Large Spin,''
JHEP {\bf 1311}, 140 (2013).
[arXiv:1212.4103 [hep-th]].
}

\lref\KomargodskiGCI{
  Z.~Komargodski, M.~Kulaxizi, A.~Parnachev and A.~Zhiboedov,
  ``Conformal Field Theories and Deep Inelastic Scattering,''
[arXiv:1601.05453 [hep-th]].
}

\lref\HellermanCBA{
  S.~Hellerman, S.~Maeda, J.~Maltz and I.~Swanson,
  ``Effective String Theory Simplified,''
JHEP {\bf 1409}, 183 (2014).
[arXiv:1405.6197 [hep-th]].
}

\lref\SiversIG{
  D.~Sivers and J.~Yellin,
  ``Review of recent work on narrow resonance models,''
Rev.\ Mod.\ Phys.\  {\bf 43}, 125 (1971).
}

\lref\PandoZayasYB{
  L.~A.~Pando Zayas, J.~Sonnenschein and D.~Vaman,
  ``Regge trajectories revisited in the gauge / string correspondence,''
Nucl.\ Phys.\ B {\bf 682}, 3 (2004).
[hep-th/0311190].
}

\lref\KarchPV{
  A.~Karch, E.~Katz, D.~T.~Son and M.~A.~Stephanov,
  ``Linear confinement and AdS/QCD,''
Phys.\ Rev.\ D {\bf 74}, 015005 (2006).
[hep-ph/0602229].
}

\lref\RychkovIQZ{
  S.~Rychkov,
  ``EPFL Lectures on Conformal Field Theory in D>= 3 Dimensions,''
[arXiv:1601.05000 [hep-th]].
}

\lref\BrowerEA{
  R.~C.~Brower, J.~Polchinski, M.~J.~Strassler and C.~I.~Tan,
  ``The Pomeron and gauge/string duality,''
JHEP {\bf 0712}, 005 (2007).
[hep-th/0603115].
}

\lref\mandelstam{
S.~Mandelstam, ``Dual-resonance models." Physics Reports 13.6 (1974): 259-353.
}

\lref\FreundHW{
  P.~G.~O.~Freund,
  ``Finite energy sum rules and bootstraps,''
Phys.\ Rev.\ Lett.\  {\bf 20}, 235 (1968).
}

\lref\MeyerJC{
  H.~B.~Meyer and M.~J.~Teper,
  ``Glueball Regge trajectories and the pomeron: A Lattice study,''
Phys.\ Lett.\ B {\bf 605}, 344 (2005).
[hep-ph/0409183].
}

\lref\CoonYW{
  D.~D.~Coon,
  ``Uniqueness of the veneziano representation,''
Phys.\ Lett.\ B {\bf 29}, 669 (1969).
}

\lref\FairlieAD{
  D.~B.~Fairlie and J.~Nuyts,
  ``A fresh look at generalized Veneziano amplitudes,''
Nucl.\ Phys.\ B {\bf 433}, 26 (1995).
[hep-th/9406043].
}

\lref\PonomarevJQK{
  D.~Ponomarev and A.~A.~Tseytlin,
  ``On quantum corrections in higher-spin theory in flat space,''
[arXiv:1603.06273 [hep-th]].
}

\lref\StromingerTalk{
  A.~Strominger, Talk at Strings 2014, Princeton.
}

\lref\CostaMG{
  M.~S.~Costa, J.~Penedones, D.~Poland and S.~Rychkov,
  ``Spinning Conformal Correlators,''
JHEP {\bf 1111}, 071 (2011).
[arXiv:1107.3554 [hep-th]].
}

\lref\CamanhoAPA{
  X.~O.~Camanho, J.~D.~Edelstein, J.~Maldacena and A.~Zhiboedov,
  ``Causality Constraints on Corrections to the Graviton Three-Point Coupling,''
JHEP {\bf 1602}, 020 (2016).
[arXiv:1407.5597 [hep-th]].
}

\lref\GrossKZA{
  D.~J.~Gross and P.~F.~Mende,
  ``The High-Energy Behavior of String Scattering Amplitudes,''
Phys.\ Lett.\ B {\bf 197}, 129 (1987).
}

\lref\KarlinerHD{
  M.~Karliner, I.~R.~Klebanov and L.~Susskind,
  ``Size and Shape of Strings,''
Int.\ J.\ Mod.\ Phys.\ A {\bf 3}, 1981 (1988).
}

\lref\SusskindAA{
  L.~Susskind,
  ``Strings, black holes and Lorentz contraction,''
Phys.\ Rev.\ D {\bf 49}, 6606 (1994).
[hep-th/9308139].
}

\lref\tHooftJZ{
  G.~'t Hooft,
  ``A Planar Diagram Theory for Strong Interactions,''
Nucl.\ Phys.\ B {\bf 72}, 461 (1974).
}

\lref\WittenKH{
  E.~Witten,
  ``Baryons in the 1/n Expansion,''
Nucl.\ Phys.\ B {\bf 160}, 57 (1979).
}

\lref\zeros{
I.~E.~Pritsker and A.~M.~Yeager,
``Zeros of Polynomials with Random Coefficients,"
J. Approx. Theory 189 (2015), 88-100.}

\lref\zerosort{
W.~V.~Assche,
``Some results on the asymptotic distribution of the zeros of orthogonal polynomials,"
Comput. Math. Appl. 12-13 (1985) 615-623.
}

\lref\Brower{
R.~C.~Brower and J.~Harte. 
``Kinematic Constraints for Infinitely Rising Regge Trajectories," 
Physical Review 164.5 (1967): 1841.
}

\lref\PolchinskiUF{
  J.~Polchinski and M.~J.~Strassler,
  ``The String dual of a confining four-dimensional gauge theory,''
[hep-th/0003136].
}

\lref\PappadopuloJK{
  D.~Pappadopulo, S.~Rychkov, J.~Espin and R.~Rattazzi,
  ``OPE Convergence in Conformal Field Theory,''
Phys.\ Rev.\ D {\bf 86}, 105043 (2012).
[arXiv:1208.6449 [hep-th]].
}

\lref\GiddingsGJ{
  S.~B.~Giddings and R.~A.~Porto,
  ``The Gravitational S-matrix,''
Phys.\ Rev.\ D {\bf 81}, 025002 (2010).
[arXiv:0908.0004 [hep-th]].
}

\lref\HellermanKBA{
  S.~Hellerman and I.~Swanson,
  ``String Theory of the Regge Intercept,''
Phys.\ Rev.\ Lett.\  {\bf 114}, no. 11, 111601 (2015).
[arXiv:1312.0999 [hep-th]].
}

\lref\AharonyIPA{
  O.~Aharony and Z.~Komargodski,
  ``The Effective Theory of Long Strings,''
JHEP {\bf 1305}, 118 (2013).
[arXiv:1302.6257 [hep-th]].
}

\lref\PolchinskiAX{
  J.~Polchinski and A.~Strominger,
  ``Effective string theory,''
Phys.\ Rev.\ Lett.\  {\bf 67}, 1681 (1991).
}

\lref\JainNZA{
  S.~Jain, M.~Mandlik, S.~Minwalla, T.~Takimi, S.~R.~Wadia and S.~Yokoyama,
 ``Unitarity, Crossing Symmetry and Duality of the S-matrix in large N Chern-Simons theories with fundamental matter,''
JHEP {\bf 1504}, 129 (2015).
[arXiv:1404.6373 [hep-th]].
}

\lref\DubovskySH{
  S.~Dubovsky, R.~Flauger and V.~Gorbenko,
  ``Effective String Theory Revisited,''
JHEP {\bf 1209}, 044 (2012).
[arXiv:1203.1054 [hep-th]].
}

\lref\LegRed{
Askey, Richard. "Orthogonal expansions with positive coefficients." Proceedings of the American Mathematical Society 16.6 (1965): 1191-1194.
}

\lref\AmatiWQ{
  D.~Amati, M.~Ciafaloni and G.~Veneziano,
  ``Superstring Collisions at Planckian Energies,''
Phys.\ Lett.\ B {\bf 197}, 81 (1987).
}

\lref\MatsudaKF{
  S.~Matsuda,
  ``Uniqueness of the veneziano representation,''
Phys.\ Rev.\  {\bf 185}, 1811 (1969).
}

\lref\KhuriAX{
  N.~N.~Khuri,
  ``Derivation of a veneziano series from the regge representation,''
Phys.\ Rev.\  {\bf 185}, 1876 (1969).
}

\lref\WeimarPI{
  E.~Weimar,
  ``Alternatives to the Veneziano Amplitude,''
    DESY-74-3.
}

\lref\WandersET{
  G.~Wanders,
  ``Constraints on the zeros and the asymptotic behavior of a veneziano amplitude,''
Phys.\ Lett.\ B {\bf 34}, 325 (1971).
}

\lref\GrossGE{
  D.~J.~Gross and J.~L.~Manes,
``The High-energy Behavior of Open String Scattering,''
Nucl.\ Phys.\ B {\bf 326}, 73 (1989).
}

\lref\ChodosGT{
  A.~Chodos and C.~B.~Thorn,
  ``Making the Massless String Massive,''
Nucl.\ Phys.\ B {\bf 72}, 509 (1974).
}

\lref\SonnenscheinPIM{
  J.~Sonnenschein,
  ``Holography Inspired Stringy Hadrons,''
[arXiv:1602.00704 [hep-th]].
}

\lref\ElShowkHT{
  S.~El-Showk, M.~F.~Paulos, D.~Poland, S.~Rychkov, D.~Simmons-Duffin and A.~Vichi,
  ``Solving the 3D Ising Model with the Conformal Bootstrap,''
Phys.\ Rev.\ D {\bf 86}, 025022 (2012).
[arXiv:1203.6064 [hep-th]].
}

\lref\KosBKA{
  F.~Kos, D.~Poland and D.~Simmons-Duffin,
  ``Bootstrapping Mixed Correlators in the 3D Ising Model,''
JHEP {\bf 1411}, 109 (2014).
[arXiv:1406.4858 [hep-th]].
}

\lref\KlebanovJA{
  I.~R.~Klebanov and A.~M.~Polyakov,
``AdS dual of the critical O(N) vector model,''
Phys.\ Lett.\ B {\bf 550}, 213 (2002).
[hep-th/0210114].
}

\lref\KorchemskyRC{
  G.~P.~Korchemsky, J.~Kotanski and A.~N.~Manashov,
``Multi-reggeon compound states and resummed anomalous dimensions in QCD,''
  Phys.\ Lett.\ B {\bf 583} (2004) 121
  doi:10.1016/j.physletb.2004.01.014
  [hep-ph/0306250].
}

\lref\NimaYutin{
N.~Arkani-Hamed, Y.~T.~Huang and T.~C.~Huang,
``String theory as the unique weakly-coupled UV Completion of YM and GR," In progress}

\lref\JoaoPedro{
M.~ Paulos, J.~Penedones, J.~Toledo, B.~C.~van Rees and P.~Vieira 
``2d S-matrix Bootstrap", To appear
}

\lref\Haus{F.~Hausdorff, ``Momentprobleme f\"ur ein endliches Intervall.'' Mathematische Zeitschrift 16.1 (1923): 220-248.}

\lref\PenedonesUE{
  J.~Penedones,
  ``Writing CFT correlation functions as AdS scattering amplitudes,''
JHEP {\bf 1103}, 025 (2011).
[arXiv:1011.1485 [hep-th]].
}

\lref\VasilievBA{
  M.~A.~Vasiliev,
``Higher spin gauge theories: Star product and AdS space,''
In *Shifman, M.A. (ed.): The many faces of the superworld* 533-610.
[hep-th/9910096].
}
\lref\MaldacenaSF{
  J.~Maldacena and A.~Zhiboedov,
  ``Constraining conformal field theories with a slightly broken higher spin symmetry,''
Class.\ Quant.\ Grav.\  {\bf 30}, 104003 (2013).
[arXiv:1204.3882 [hep-th]].
}

\lref\Tao{
T.~Tao, V.~Vu. 
``Local universality of zeroes of random polynomials." 
International Mathematics Research Notices (2014): rnu084,
[arXiv:1307.4357 [math.PR]].
}

\lref\MakeenkoRF{
  Y.~Makeenko and P.~Olesen,
  ``Wilson Loops and QCD/String Scattering Amplitudes,''
Phys.\ Rev.\ D {\bf 80}, 026002 (2009).
[arXiv:0903.4114 [hep-th]].
}

\lref\ArmoniNJA{
  A.~Armoni,
  ``Large-N QCD and the Veneziano Amplitude,''
Phys.\ Lett.\ B {\bf 756}, 328 (2016).
[arXiv:1509.03077 [hep-th]].
}

\lref\ArkaniHamedBZA{
  N.~Arkani-Hamed and J.~Maldacena,
  ``Cosmological Collider Physics,''
[arXiv:1503.08043 [hep-th]].
}

\Title{
\vbox{\baselineskip6pt
}}
{\vbox{
\centerline{Strings from Massive Higher Spins:}
\vskip 0.15in
\centerline{The Asymptotic Uniqueness of the Veneziano Amplitude}
}}

\bigskip
\centerline{Simon Caron-Huot,$^{1}$ Zohar Komargodski,$^{2}$ Amit Sever,$^{3}$ and Alexander Zhiboedov$^{4}$}
\bigskip
\centerline{\it $^{1}$ Niels Bohr International Academy and Discovery Center, }
\centerline{\it Blegdamsvej 17, Copenhagen 2100, Denmark}
\centerline{\it $^{2}$ Weizmann Institute of Science, Rehovot 76100, Israel}
\centerline{\it $^{3}$ School of Physics and Astronomy, Tel Aviv University, Ramat Aviv 69978, Israel}
\centerline{\it $^{4}$ Department of Physics, Harvard University, Cambridge, MA 20138, USA
}

\vskip .2in 

\noindent
We consider weakly coupled theories of massive higher-spin particles. This class of models includes, for instance, tree-level String Theory and Large-N Yang-Mills theory. The S-matrix in such theories  is a meromorphic function obeying unitarity and crossing symmetry. 
We discuss the (unphysical) regime $s,t \gg 1$, in which we expect the amplitude to be universal and exponentially large. We develop methods to study this regime and show that the amplitude necessarily coincides with the Veneziano amplitude there. In particular, this implies that the leading Regge trajectory, $j(t)$, is asymptotically linear in Yang-Mills theory. Further, our analysis shows that any such theory of higher-spin particles has stringy excitations and infinitely many asymptotically parallel subleading trajectories. More generally, we argue that, under some assumptions, any theory with at least one 
higher-spin particle must have strings.  

\Date{ }

\listtoc\writetoc
\vskip .5in \noindent

\newsec{Introduction}

In this note we consider theories that contain massive particles of arbitrarily high spin. We assume that the particles interact weakly and are thus approximately stable. This situation arises in several cases. First, tree-level string theory consists of infinitely many spinning particles. At the tree approximation these particles are exactly stable~\VenezianoYB. Second, we can study large-N Yang-Mills theory (such as pure Yang-Mills theory or other confining gauge theories). At large N the theory can be described by approximately stable resonances (glueballs) with various masses and spins~\refs{\tHooftJZ,\WittenKH}. While at first sight Yang-Mills theory is different from classical string theory, Yang-Mills theory does lead to stringy excitations. The color flux lines are confined to flux tubes of finite width but arbitrary length. 

More generally, one may wonder what are the possible theories of weakly interacting massive higher-spin particles. Here our goal is to show that any such theory must contain strings.

Theories of weakly interacting higher-spin particles are strongly constrained by various consistency conditions. In this note we will mostly restrict our attention to the consistency conditions that follow
from the S-matrix of such theories.\foot{We review some basic ideas in the subject of scattering amplitudes in appendix~A.1.} 

Let us imagine that our theory consists of the fields $\phi^{(n)}_{\mu_1...\mu_L}$ with mass squared $m^2_{n,L}$. We imagine a $2\to2$ scattering process of the scalar particle $S$. In general, there would be contributions from Feynman diagrams in the $s$, $t$ and $u$ channels. The $s$-channel contributions lead to a delta-function discontinuity in the variable $s$ at the location of the resonance. The same is true for $u$-channel diagrams but not for $t$-channel diagrams ($t$-channel diagrams do not lead to discontinuities in the variable $s$). To make the discussion a little simpler, in the rest of the paper, we assume that the $u$-channel resonances are absent. This can be arranged by scattering non-identical particles, such that there are no resonances with the right quantum numbers in the $u$-channel. However, we can retain the symmetry between $s$ and $t$ channel processes. This simplification is only for technical reasons and our arguments can be easily generalized to incorporate the $u$-channel resonances.

Under these assumptions, the imaginary part of the amplitude in the variable $s$ is localized to the $s$-channel poles
\eqn\onereii{ {\rm Im}_s[ A(s,t) ]=\sum_{n,L} f_{n,L}^2\delta\left(s-m^2_{n,L}\right)P_L\left(1+{2t\over m_{n,L}^2-4m_S^2}\right)~, }
where $P_L(x)$ are the $D$-dimensional cousins of Legendre polynomials; see appendix A.1 for details. 
The $f^2_{n,L}$ are positive coefficients, related to the strength of the coupling of $S$ to the intermediate states.

Without further restrictions, the scattering amplitude with the properties above is not so tightly constrained. In other words,~\onereii\ and $A(s,t)=A(t,s)$ allow for many different solutions, many of which are uninteresting to us (in particular, any classical field theory would solve these constraints). 

 To make further progress we impose restrictions on the high energy behavior of the amplitude.
The condition that makes the problem nontrivial (and appropriate for studying theories of infinitely many higher-spin particles) is the following. Given that there is a particle of spin $L$ in the spectrum, we assume that there exists a $t_0$ such that
\eqn\highenergybound{
\lim_{|s| \to \infty} s^{-L} A(s, t_0) = 0~.
}
This does not have to be true for all $L$, but only for some $L$ in the spectrum of particles.

One way to motivate this condition is to imagine that we have a massive higher-spin particle of $L>2$ being exchanged and impose the causality constraint of~\CamanhoAPA. As argued in~\CamanhoAPA, massive higher-spin particles naturally appear in classical gravitational theories with higher derivative corrections. Then causality implies that~\highenergybound\ holds in these theories. Another way to motivate~\highenergybound\ is to note that any theory with an ultra-violet fixed point is expected to behave in a power law fashion in the hard scattering regime. Therefore, in Yang-Mills theory it would be sufficient to choose some large negative $t_0$ for~\highenergybound\ to be satisfied for some $L$. More generally, we can say that polynomial boundedness is a general property of QFT (see~\GiddingsGJ\ for a recent discussion and references).

For example, setting in the formula above $L=0$, if for some range of $t$ we have that 
\eqn\convrange{\lim_{|s|\rightarrow\infty}A(s,t)\rightarrow 0~,} then we can use the unsubtracted dispersion relation 
\eqn\dispersion{
A(s,t)=\int_{0}^\infty ds' { {\rm Im}_{s'}[ A(s',t)  ] \over s'-s}~
}
in that range of $t$. Typically this range would be defined by $t<t_0$ for some $t_0$ (we will return to the question of whether $t_0>-\infty$ soon).

Plugging~\onereii\ into~\dispersion\ we find 
\eqn\ampartial{A(s,t)=\sum_{n,L}f_{n,L}^2{P_L\left(1+{2t\over m_{n,L}^2-4 m_S^2}\right)\over s-m^2_{n,L}}~.}
This representation of the amplitude is similar, but not identical to the standard partial wave expansion. 

Since the amplitude satisfies crossing symmetry (which amounts to exchanging identical particles and is manifest in the description in terms of Feynman diagrams)
\eqn\crossing{A(s,t)=A(t,s)}
one has to require this property from the decomposition~\ampartial.
This is a highly nontrivial constraint on the spectrum of masses $m_{n,L}^2$ and on the coefficients $f_{n,L}^2$. (Of course, $m_S^2$ is just one of the particles
labeled by $n,L$ with $L=0$.)

Assuming that~\convrange\ is satisfied in some range of $t$, one can immediately conclude from~\ampartial\ that the theory must have infinitely many resonances. This is because if the number of resonances is finite the amplitude is polynomial in $t$ which contradicts crossing symmetry~\crossing\ (from which it follows that there must be poles in $t$ at the same places where there are poles in $s$). 

Further, one can conclude immediately that there must be some resonances with spin $L>L_0$ for any $L_0$. In other words, the spin of the resonances must be unbounded from above. This is again because if there is an upper limit on the spin, $L_0$, taking $L_0+1$ derivatives with respect to $t$ of~\ampartial\ would annihilate the amplitude, contradicting~\crossing. 

We reach these conclusions about having infinitely many particles with unbounded spin if we assume~\highenergybound. As discussed after~\onereii, not all the consistent $S$-matrices are of this type, for example, classical field theories can provide counter-examples. We could write
$$A(s,t)\sim {1\over s-m_\phi^2}+{1\over t-m^2_\phi}~,$$
which is the tree-level amplitude of one stable scalar particle with a cubic interaction vertex $\phi^3$. Here we clearly see that~\convrange\ (and~\highenergybound) are not satisfied for any $t$ and thus even though~\onereii\ still holds true, the conclusion about infinitely many particles with unbounded spin does not follow.\foot{Our discussion also does not apply to the $O(N)$ models and related theories because the $S$-matrix is not meromorphic even at infinite $N$. See~\JainNZA, where the $S$-matrix for the elementary fields is discussed. } The condition~\highenergybound\ is therefore interesting. It allows to study theories with higher-spin particles, excluding all the classical field theories with finitely many fields, but retaining Yang-Mills-like theories and String Theory. (Also theories such as the Large N ${\cal{N}}=1^*$ theory~\refs{\PolchinskiUF}, which has an interacting ultraviolet fixed point, satisfy~\highenergybound.)

Let us now fix some real $t$ and take $|s| \gg |t|,m^2_S$. We can parametrize the behaviour of $A(s,t)$ in this limit by\foot{Some properties of $F(t)$ are reviewed in appendix A.2.}
\eqn\definitionRegge{
\lim_{|s| \gg |t|,m^2_S} A(s,t) = F(t) (-s)^{j(t)}\ ,\qquad{\rm arg}[s] \neq 0 \ .
}
One has to be slightly careful in taking the limit as in~\definitionRegge. Along the positive real axis, there are poles when $s$ hits resonances. Therefore, one should excise some infinitesimal wedge around the positive real axis for the limit~\definitionRegge\ to exist. The minus sign in front of $s$ indicates that the $s$-cut is along the positive real axis where the poles are distributed.

In the region of $t<0$, the limit of $s\gg |t|$ describes  physical small angle scattering. It is known as the Regge limit. If in some region $j(t)<0$ (typically this happens at negative $t$) then the condition~\convrange\ is satisfied and we can use the representation~\ampartial.\foot{In tree-level string theory, $j(t)=\alpha't+\alpha_0$.} More generally, if the theory has infinitely many higher-spin particles and if for negative $t$ the function $j(t)$ is bounded from above, the representation~\highenergybound\ applies (this is satisfied in any QFT due to the polynomial boundedness property that we alluded to above).

The interpretation in the regime $t>0$ is quite different. The points $t=t_n$ where \eqn\leadingtra{j(t_n)=n} with $n$ a non-negative integer are special because the dependence on $s$ is a polynomial. This is exactly what we would expect at the points where $t$ hits a resonance. In appendix A.2 we present a simple argument that it is indeed the case. Therefore, the $t_n$, which are the solutions of~\leadingtra, are identified with resonances of mass squared $m^2=t_n$ and spin $n$. These are the fastest spinning particles in the theory, since they are the dominant ones at large $s$ when $t$ is near a pole. Hence, the solutions to~\leadingtra\ define the particles on the leading Regge trajectory.

While there is a very simple formula for $j(t)$ in tree-level string theory, much less is known about Large-N Yang-Mills theory.
There is no reason to expect that the leading trajectory is exactly linear. In fact, consider the high-energy fixed-angle scattering regime, where $-t,s\gg \Lambda^2$ and $\Lambda$ is the strong-coupling scale. In that limit, the linear behaviour predicts an exponentially small scattering amplitude. This is the famous soft behaviour of strings at high energies \GrossKZA . But in QFT one can at most get power laws and logarithms.\foot{In this paper a ``QFT"
refers to any theory which approaches a scale-invariant fixed point in the UV. This includes asymptotically free gauge theories but also more general setups.}
Therefore, one expects that in QFT  $$\lim_{t\rightarrow-\infty}j(t)\rightarrow {\rm const}$$
up to logarithmic corrections.

\ifig\figone{Expected shape of the leading Regge trajectory in QFT. At large positive $t$ it is linear while at large negative $t$ it approaches a constant. (As far as we know, the transition could be smooth or first order. Here, the origin of the axes is arbitrary.)} {\epsfxsize 3.3in \epsfbox{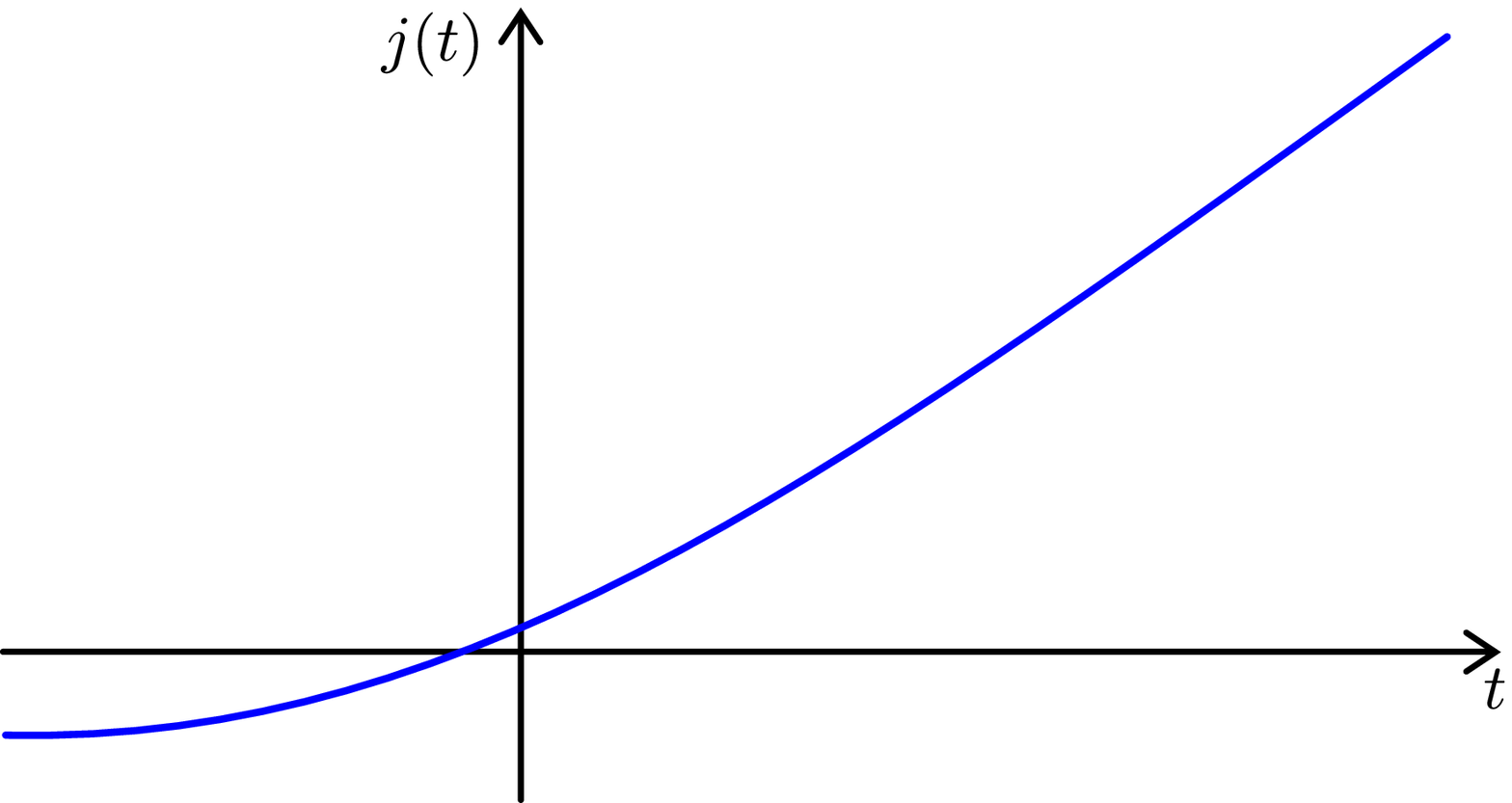}}

This suggests that the behaviour of $j(t)$ for large negative $t$ (fixed angle scattering) or finite negative $t$ (small angle scattering) is not universal and may depend on the theory that is being studied. However, the behaviour for large positive $t$ may well be universal. Large positive $t,s$ does not correspond to 
physical $2\to2$ scattering (as the scattering angle in this regime is imaginary) but the structure of $j(t)$ at large $t$ allows to learn something about the structure of heavy spinning resonances~\leadingtra\ and it also allows to access the scattering amplitude in the high-energy large impact parameter regime.   The result in the large impact parameter regime-- see appendix A.3 for details -- is 
\eqn\decayingamp{
{\rm Im}\,A(b,s)\sim e^{-{b^2\over 4\alpha' \log(s)}}~.
}
This is the dominant contribution to the {\it inelastic} part of the amplitude.

The exponential decay in~\decayingamp\ is very reasonable since at impact parameters larger than the natural scale in the problem, one expects very little scattering, not to mention creating on-shell intermediate configurations.  The fact that the inelastic amplitude is nonzero at all at finite impact parameter and tree-level is indicative of the presence of extended objects in the theory.
In the context of string theory, the fact that the effective distance is logarithmically enhanced by $\log(s)$ is also indicative of the fluctuations of the string. Indeed, the quantization of the string leads to $\langle X_\perp^2\rangle \sim \log(s)$ for the orthogonal length of the string~\refs{\KarlinerHD, \SusskindAA}. We therefore see that a behaviour of the form $j(t)\sim \alpha't$ at large positive $t$ is indicative of the existence of a string.

In summary, we would expect that in QFT $j(t)$ would take the schematic form in Fig.1. In fact, this is roughly how the Regge trajectory looks like in 
the context of the gauge-gravity duality~\BrowerEA. At large positive $s,t$, which is the regime we study in this paper, the amplitude becomes exponentially large. This is perhaps related to the fact that this regime is universal. 

In this note, our purpose is to establish the above-claimed asymptotic linearity of $j(t)$
\eqn\asymlin{
j(t) = \alpha' t + ...\ ,\qquad t \gg 1 \ .
}
We will see that this follows from the highly nontrivial consistency conditions on scattering
amplitudes that we explained above. In fact, we do slightly more: we show that in the unphysical region of large positive $s,t$, the consistency conditions force the amplitude to coincide with the Veneziano amplitude
\eqn\uniqueampl{
\lim_{s,t \gg 1} \log A(s,t) = \alpha'\Big((s+t) \log (s+t) - s \log s - t \log t\Big) \ .
}

Therefore, there is a limit where the scattering amplitude of any theory of weakly interacting massive higher-spin particles (including large N Yang-Mills theory) must coincide with the stringy amplitude.

According to the discussion above, this implies $m^2\sim L$ for the leading Regge trajectory at large masses. Additionally, it implies a stringy picture for the bound states. Therefore, in some sense we can prove that there are strings in Large-N Yang-Mills theory.\foot{In the physical regime and for large momentum transfer, an attempt to connect the Veneziano amplitude to scattering in large-N QCD  was made in \refs{\MakeenkoRF,\ArmoniNJA}. However, the examples studied in \BrowerEA\ show that this regime is possibly not universal.}

Let us also emphasize that we assume throughout our paper that the there is no accumulation point in the spectrum, namely that for any $M$ there is a finite number of particles with mass $m \leq M$. Relaxing this assumption leads to possibilities different from~\asymlin. Theories that lead to such amplitudes are non-local at some scale below $1/M$.  As a simple example imagine that $m_S^2 =0$ and consider the following amplitude\foot{ This example is somewhat reminiscent of the 3D vector $O(N)$ model which is dual to the Vasiliev theory in AdS~\refs{\KlebanovJA,\VasilievBA}. In this case there is only one Regge trajectory but this happens at the cost of locality. Here the loss of locality manifests itself through the presence of additional singularities. }
\eqn\toysolution{
A(s,t) \sim {1 \over (s-m_{\phi}^2) (t-m_{\phi}^2)} \ .
}
This amplitude is crossing symmetric and contains an infinite number of massive higher spin particles with mass $m_{\phi}$. One can check that it is unitary in the sense that all the coefficients of the Legendre polynomials in~\onereii\ are non-negative. Another amplitude with an accumulation point in the spectrum was found by Coon~\CoonYW. In this case one finds logarithmic trajectories $j(t) = \log t +...$ (see appendix C).

This problem of determining $j(t)$ has a very long history that we will not attempt to cover fully.\foot{There was some work in the past on the uniqueness of the Veneziano amplitude, see e.g.~\refs{\MatsudaKF\KhuriAX-\WeimarPI}. However, the Regge trajectories were assumed to be exactly linear. We do not make this assumption. For example, in QFT the Regge trajectories are not expected to be exactly linear. That would have been in contradiction with the existence of an ultraviolet fixed point.   Our goal is to derive asymptotic linearity by making the appropriate assumptions.} One notable development is the so-called ``Mandelstam argument''~\mandelstam , see appendix A.4 where the argument is reviewed. 

Another approach assumes that the theory admits extended flux tubes. One then quantizes this theory (see~\refs{\DubovskySH,\PolchinskiAX,\AharonyIPA,\HellermanCBA} and references therein) and attempts to study fast spinning flux tube configurations~\HellermanKBA. So far this approach has had limited success due to the cusp singularities that develop on spinning closed strings. This, however, may be just a technical problem and once we learn how to renormalize these singularities progress could be made. Finally, there is also a large amount of literature about the subject in the context of holography. See for instance~\refs{\PandoZayasYB,\KarchPV}.

Of course, the approximate linearity of trajectories in Yang-Mills theory is observed in nature as well as on the lattice 
(for some lattice results see for instance~\MeyerJC). However, the resonances become exactly stable only at large $N$ and this is where we attempt a rigorous 
understanding of this phenomenon.  

It would be interesting to push our methods further and compute the corrections in~\asymlin. This can be compared with the various other methods we alluded to above. We leave that to future work.

We now summarize our assumptions and briefly explain the methods we will use in order to derive from them the asymptotic form of the amplitude~\uniqueampl. In this sketch we only stress the main ideas and avoid rigour. 

\subsec{Sketch of the Paper}

 We work under the following assumptions elucidated above

\item{1)} {\bf Tree-level/weak coupling:} $A(s,t)$ is a meromorphic function with only simple poles \onereii\ at the location of resonances.
\item{2)} {\bf Unitarity:} Residues are sums of Legendre polynomials with non-negative coefficients \onereii.
\item{3)} {\bf Crossing:} $A(s,t)=A(t,s)$.
\item{4)} {\bf High energy behaviour:} There exists a particle of spin $L$ in the spectrum and some $t_0$ such that \highenergybound\ holds.
\item{5)} {\bf No accumulation point in the spectrum:} For any $M$, the number of particles with the mass $m<M$ is finite.
\item{6)} {\bf Asymptotic Regge limit:} We assume that the usual Regge limit asymptotic formula~\definitionRegge\ controls not only  the $t$ - fixed, $s \to \infty$ limit, but ${t \over s} \ll 1$ - fixed and $s,t \to \infty$ limit as well. 

\medskip

Let us make an additional comment regarding $6)$. In physical terms we assume that in the large $s,t \gg 1$ region all intermediate scales decouple, or, in other words, threshold effects do not persist, and we can smoothly interpolate between the usual Regge limit for $t$ being large but fixed and the asymptotic Regge limit described above. 

The main idea will be to study the zeros of the amplitude.  Consider the amplitude $A(s,t)$ now as a function of $s$ for some fixed $t$. It is a  meromorphic function with poles only on the positive axis at 
$s=m_{n,L}^2$. At each such pole the residue is a sum of Legendre polynomials with non-negative coefficients.
$$\lim_{s\rightarrow m^2}A(s,t)\quad\longrightarrow\quad{1\over s-m^2}\sum_{n,L}f_{n,L}^2P_L\left(1+{2t\over m^2-4m_S^2}\right)~,$$
where the sum runs over all the particles of mass $m^2$. Since Legendre polynomials $P_L(x)$ are positive for $x>1$, we see that the residue is positive for
positive $t$ and for $m^2>4m_S^2$. Therefore, the amplitude $A(s,t)$ has various poles, but for large enough $s$, all the residues are positive for $t>0$.
{\it This means that between any two adjacent poles there must be at least one zero.} This assertion crucially depends on unitarity.  The amplitude may have additional zeros, which are not between two poles. See Fig 2. for a schematic depiction of the amplitude for some positive values of $t$. 

For large enough $s$, between every two poles there will be exactly one zero. Otherwise, the Regge asymptotic \definitionRegge\ will not hold true.  This can be shown as follows: If we consider $\log A(s,t)$
then for large enough $s$ we can use~\definitionRegge, which gives $\log A(s,t)=j(t)\log(-s) +...$. The total discontinuity in $s$ is therefore $j(t)$. But from the zeros and poles of the amplitude, we get that every pole contributes $-1$ to the discontinuity in $\log A$ and every zero contributes $+1$. This can be seen from 
\eqn\discamp{
{\rm Disc}[ \log A ] =\int dz \ {\pa_s A \over A} ~,
}
which receives contributions from poles and zeros in $s$, with opposite signs.

Therefore 
\eqn\zerospoles{\sum\left(\#{\rm zeros}(t)-\# {\rm poles} \right)=j(t)~.}
Of course this formula requires some sort of regularization, which we will discuss in the body of the paper.

As we vary $t$, the zeros move (the poles are at fixed locations). When $t$ hits a resonance, there is a zero in $s$ at exactly the points where there is a pole in $s$ so they cancel each other. This is necessary in order to obtain a polynomial in $s$ at the resonances in $t$. See Fig. 2.

\ifig\figone{Poles in $s$ are represented by red boxes and zeros by blue crosses. The poles are fixed at the resonances while the zeros can move. At positive $t$ we have one zero between any two successive poles (unitary zeros) plus a finite set of so-called excess zeros. As we increase $t$, the unitary zeros flow to the left. When $t$ hits a resonance, there is a zero in $s$ at exactly the points where there is a pole in $s$, so that they cancel each other and we remain with a polynomial. As $t$ cross $m_i^2$, one unitary zero becomes an excess zero and so the number of excess zeros has grown by one, in agreement with~\zerospoles.} {\epsfxsize 5.2in \epsfbox{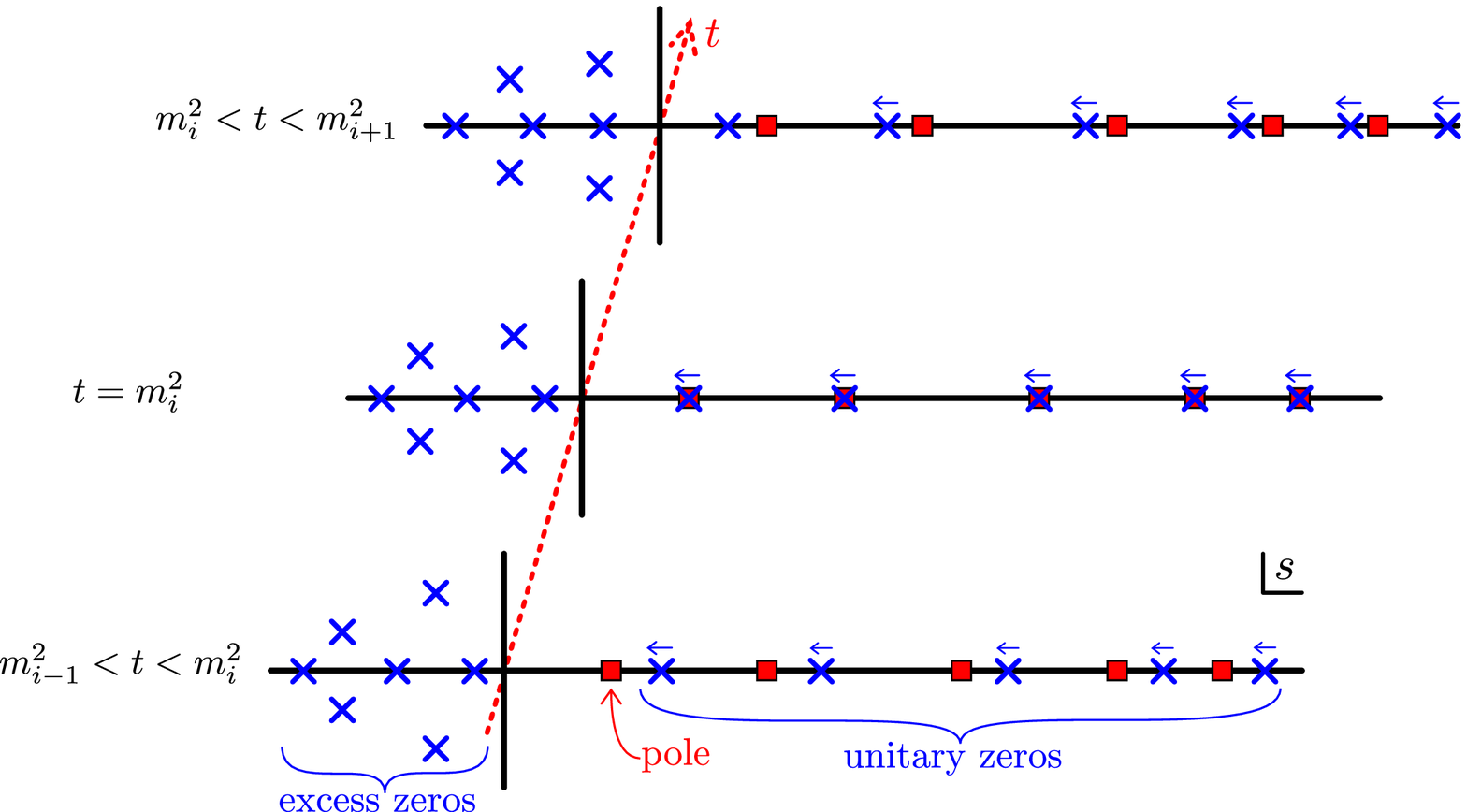}}

It is thus clear from~\zerospoles\ that to learn about $j(t)$ at large positive $t$ we need only to concentrate on the excess zeros. There are increasingly more 
excess zeros as we increase $t$ along the positive axis. At $t=m_{n,L}^2$ the excess zeros are those of positive sums of Legendre polynomials. As we 
increase $t$ from one pole on the leading trajectory to the next, those should be Legendre polynomials of increasing rank. 

The next crucial step is that we introduce a ``thermodynamic'' approach to study these excess zeros. We assume that they form a distribution. This distribution can be viewed as coarse graining over the Legendre polynomials that appear in nearby resonances. It turns out that, under mild assumptions, the distribution of zeros of positive sums over Legendre polynomials has support inside the unit disc. We show that this property together with unitarity lead to only one possible solution, namely,~\uniqueampl.

Therefore, in this sense, we derive the stringy picture and the asymptotic linearity of the leading trajectory in any theory with weakly interacting higher-spin particles. 

The paper is organized as follows. In section~2 we discuss a convenient representation of the scattering amplitudes using infinite products. This makes the dynamics of zeros more transparent. 
In section~3 we discuss the distribution of zeros and its relation to the amplitude for large, positive, $s,t$.
In section~4 we study the consequences of unitarity for the distribution of zeros. 
In section~5 we study in more detail the consequences of crossing symmetry and analyticity. We arrive at the conclusion that the only consistent asymptotic form is~\uniqueampl. Finally, we conclude in section 7 and discuss various open problems.

In appendix A~we collect some basics facts about scattering amplitudes. In appendix~B we prove that large-N Yang-Mills theory must have infinitely many spin 0 and spin 2 resonances using simple QFT arguments. In appendix~C we discuss  various other (unphysical) possible solutions to some of our constraints. In appendix~D we review the salient features of the Veneziano amplitude.

\newsec{Product Representation}

If we have a meromorphic function $f(z)$ with zeros at $\{z_i\}$, simple poles at $\{p_i\}$, and no other singularities, and if we assume 
\eqn\convergencei{\sum_i \left|{1\over z_i}-{1\over p_i}\right| <\infty~, }
then the function can be represented in the Weierstrass product form as\foot{Otherwise, one needs to include elementary factors for the product representation to converge.}
\eqn\Wform{f(z)=z^m e^{g(z)}\prod_i \left(     {1-{z\over z_i}\over 1-{z\over p_i}}   \right)}
with $g(z)$ an entire function. 

In particular, let us assume that, asymptotically, the poles scale as $p_i\sim i^k$ ($k>0$) and that the zeros are in between poles. (There may be finitely many zeros which are not of this form.)
Then,  $ \sum_i \left|{1\over z_i}-{1\over p_i}\right| \sim\sum_i {i^{k-1}\over i^{2k}}\sim \sum_i {1\over i^{k+1}}  $, which converges for positive $k$.
Under the same assumptions, if we require that away from the positive axis the function $f(z)$ behaves like $z^j$ at $|z|\rightarrow\infty$, then
we can can set $e^{g(z)}=const$.

We have to be careful before we conclude that therefore the representation of the form~\Wform\ is applicable for the scattering amplitude. This is because the scaling ansatz $p_i\sim i^k$ postulated above is strictly speaking only applicable to the scattering amplitudes with the property that all its poles belong to the leading Regge trajectory. In the general case we still have a zero between every two poles but apart from the poles of the leading Regge trajectory we also have poles that correspond to the subleading Regge trajectories. Let us estimate how the presence of the subleading trajectories modifies the  convergence argument above. 

We can combine all the zeros between every two poles of the leading Regge trajectory as well as the subleading poles to get
\eqn\subleadingdifference{
  \sum_{k=1}^{k_{max}(i)} \left| {1 \over z_{i ,k} } - {1 \over p_{i,k} } \right| \leq {1 \over i^k} - {1 \over (i+1)^k} \ ,
}
where the upper estimate comes from moving all the zeros to the right until they hit the next pole. Thus, the previous argument applies and the product converges.

Therefore in the context of scattering amplitudes we can write\foot{For an early discussion of the product representation in the context of scattering amplitudes see, for example, \WandersET. } 
\eqn\product{
A(s,t)=F(t) \prod_i \left(     {1-{s\over z_i(t)}\over 1-{s\over m^2_i}}   \right)
}
where the index $i$ runs over all the resonances, and in order for the notation to remain simple, we label the zeros with the same index as the poles since asymptotically there is a zero between every two resonances for positive $t$. Crossing symmetry is highly nontrivial in this product representation. This will be our main topic of discussion in the next sections. 
Let us mention a straightforward property of the product representation~\product:
 If we take $s\sim m_i^2$ the residue must be given by a polynomial in $t$. By duality, the same must be true for $t\sim m_i^2$. So each and every pole in $s$ in~\product\ has to cancel out when we plug $t=m_i^2$. This means that for every $i$ and every $k$ there is a $j$ such that $z_j(m_k^2)=m_i^2$.

\newsec{Asymptotic Distribution of Zeros}

As explained above, a meromorphic amplitude in a unitary theory admits the product representation of the form
\eqn\productreps{
A(s,t) = F(t)  \prod_{i = 1}^{\infty} \left(     {1-{s\over z_i (t)}\over 1-{s\over m^2_i}}   \right)  \ .
}
As we have already explained in detail, in unitary amplitudes, for positive $t$, between any two poles there should be at least one zero. Furthermore, Regge behavior implies that at large enough $s$ there is exactly one zero between two adjacent poles. We refer to these zeros as unitary zeros. Let us therefore focus on the excess zeros. They are not paired up with poles, their number depends on $t$ (there is a finite number of them for finite $t$) and they dominate the Regge behavior (i.e. the integer part of $j(t)$)~\definitionRegge,~\zerospoles,
 as well as the logarithm of the amplitude for $s,t$ large.

The zeros that are paired with poles make a small contribution in the Regge limit. Indeed, at large $s$ a rational function $s-z(t)\over s-m^2$ behaves like $1+\CO(1/s)$ and as a result one can bound the contribution from the paired zeros to be smaller or comparable to $O(1)$. 

Let us label these excess zeros by $z^e (t)$. Below we will study these excess zeros in detail and see that crossing symmetry $A(s,t) = A(t,s)$ leads to nontrivial restrictions.
Indeed, since in the large $s,t$ limit unitary zeros and poles screen each other, the amplitude is dominated by the excess zeros. The excess zeros are then constrained by crossing symmetry, which maps the region of large $s,t$ to itself.\foot{If we have had the $u$-channel, there would have been another set of poles and unitary zeros that screen each other and do not have any effect on our discussion.} 

Let us start by writing $\log A(s,t)$ in terms of the distributions of the excess zeros
  \eqn\logA{\eqalign{
 \log A(s,t) &= \log F(t) + \int d^2 z \ \rho(t,z, \bar z) \left[\log \left( z - s \right)-\log(z)\right] + ... \ , \cr
 \rho(t,z, \bar z) &= \sum_{i} \delta^{(2)}(z - z_{i}^e(t)  ) \geq 0 \ .
 }}
The corrections in~$\cdots$ above are sub-dominant in the large $s,t$ limit.\foot{We have written $\log \left( z - s \right)-\log(z)$ instead of $\log \left( 1 - {s\over z} \right)$ to emphasize that the branch cut in $s$ is at positive $s$. Below we are less explicit 
about this, but it should be kept it mind that the discontinuity of the amplitude is at positive $s$. }
The leading piece of the amplitude, controlled by the distribution of the excess zeros $\rho(t, z, \bar z)$, as given by~\logA, should be crossing symmetric up to corrections which are small in the large $s,t$ limit. This condition constrains $\rho(t, z, \bar z)$.

\ifig\figdistas{The large $s$ asymptotic region where we defined the asymptotic distribution of excess zeros $\rho$.} {\epsfxsize 3.5in \epsfbox{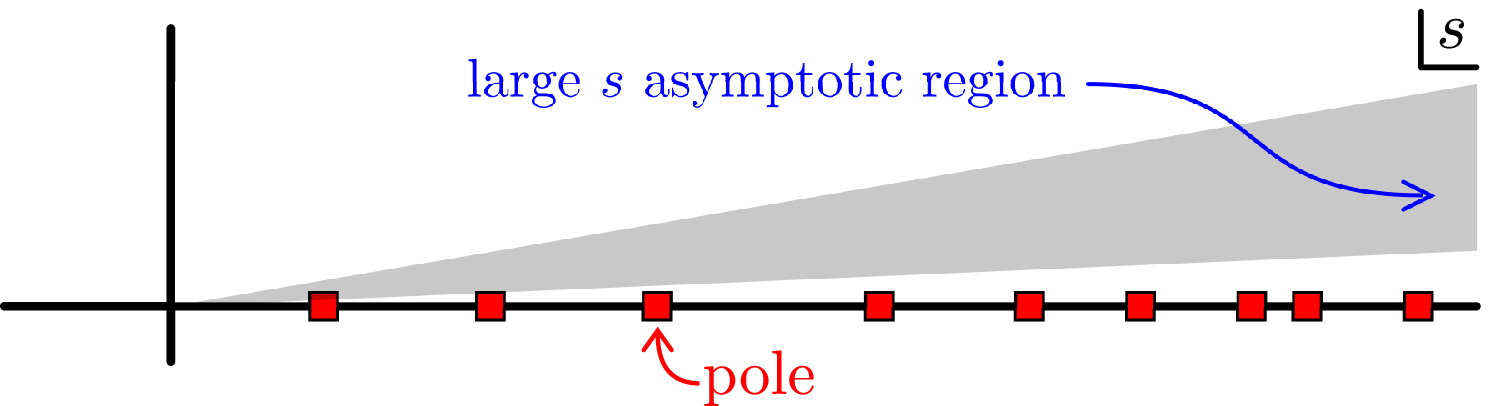}}

Now we make a crucial step, which we can justify only heuristically. We assume that at large $t$ $\rho(t, z, \bar z)$ is given by a smooth  distribution.
This is motivated in the following way: At large $t$ the number of zeros is very large. If we study the amplitude not on the real $s$ axis but at some angle $\pi-\delta$ with $|\delta|\ll 1$ then the amplitude itself does not have poles or zeros as those are averaged over, see figure 3. At large $s,t$ the averaging is over increasingly many zeros and poles since the distance from the real axis is large.\foot{A convenient way to perform this average is to consider $A(s(1+ i \eps), t(1 + i \eps))$. See appendix~D, where it is explicitly demonstrated in the case of the Veneziano amplitude.} 
The logarithm of the amplitude behaves more smoothly and we can view~\logA\ as the definition of the distribution that one obtains. 
We will soon see that this assumption that $\rho(t,z,\bar z)$ behaves like a smooth distribution is not very restrictive by itself as many possible distributions exist, corresponding to various possible forms for the Regge trajectories. 

When $t$ is large the dynamics of the excess zeros is controlled by $t$ and hence dimensional analysis restricts us to the following ansatz for $\rho(t,z, \bar z)$ (this assumption means that threshold effects of the low-mass particles disappear in the asymptotic regime of large $s,t$)
\eqn\ansatzdensity{
\rho(t, z, \bar z) = {j(t) \over t^2} \rho\left({z \over t},{\bar z \over t}\right) + ...\ ,\qquad t \gg 1 . 
}
The prefactor $j(t)$ is dictated by the Regge limit~\definitionRegge. By writing the distribution of zeros in this way, $\rho$ is now a normalized distribution with 
\eqn\norma{\int d^2 z \ \rho(z, \bar z) = 1~.}
In order to verify~\norma, we plug \ansatzdensity\ into \logA\ and rescale the integration variable as $z \to t z$. We get the following expression for the amplitude
\eqn\logAb{
\log A(s,t) = \log F(t) + j(t) \int d^2 z \ \rho(z, \bar z) \log \left( 1 - {s \over t z} \right) \ .
}
This form makes the behavior in the Regge limit $s\gg t$ manifest. Indeed we get that
\eqn\ReggelimitS{
\lim_{s \gg t} \log A(s,t)  = j(t) \log(-s) \ \int d^2 z \ \rho(z, \bar z) =j(t) \log(-s)~.
}
We have used~\norma. The result~\ReggelimitS\ is consistent with the amplitude being characterized in the asymptotic region by the positive-definite density of zeros $\rho(z, \bar z)$ normalized to $1$ as above.

\subsec{Reproducing the Regge Limit in the Dual Channel}

It is very instructive to use \logAb\ to reproduce the behavior in the dual Regge limit $t \gg s$. The crucial observation is that in \logAb\ the nontrivial dependence on $s$ enters only through the ratio ${s \over t}$, namely we have
\eqn\simpleobservation{\eqalign{
\log A(s,t) &= \log F(t) + j(t) a(\beta)\ ,\qquad\beta = {s \over t}  \ , \cr
a (\beta) &= \int d^2 z \ \rho(z, \bar z) \log \left( 1 - {s \over t z} \right)  \ . 
}}
In the Regge limit $t \gg s$ we, thus, get
$$\log F(t) + j(t) a(\beta)=j(s)\log(-t)~,$$
i.e.,
\eqn\ReggelimitDual{
\lim_{t \gg s}  {j(s) \log(-t) - \log F(t) \over j(t) }  = a(\beta) ~.
}
Above we have neglected subleading terms in the large $s,t$ limit.
An example of an interesting solution to this equation is \eqn\uniquesolution{\eqalign{
j(t) &= t^k\ ,\qquad\log F(t) = 0 \ .
}}
This describes Regge trajectories which are asymptotically a power $t^k$. Constraining the admissible values of $k$ and showing that only $k=1$ is consistent with unitarity and crossing will be the main subject of the rest of the paper.
The equation~\ReggelimitDual\ admits additional solutions which describe various less relevant situations. There are also situations where~\ansatzdensity\ is not satisfied because the asymptotic distribution depends on threshold scales. We discuss some of these cases in appendix~B.

\subsec{Crossing Equation}

Using the solution \uniquesolution\ we get the following expression for the scattering amplitude
\eqn\distrcomp{
\log A(s,t) =  \int d^2 z\,  \rho(z, \bar z)\, t^{k} \log (1-{s\over t z})\ ,\qquad \rho(z,\bar z) \geq 0 \ .
}
Notice that this expression exhibits a scaling symmetry
\eqn\scalingsymmetry{
\log A(\lambda s, \lambda t) = \lambda^{k} \log A(s,t) .
}

After this long preparation we can finally write the crossing equation for the distribution $\rho$: 
\eqn\crossingequation{\eqalign{
&\int d^2 z\,  \rho(z, \bar z)\, \left( \beta^{k} \log (1-{1\over \beta z})  - \log (1-{\beta \over z})  \right) =0\ ,\qquad \beta = {s \over t} > 0~, \ \cr
&\int d^2 z\,  \rho(z, \bar z) = 1\ , \qquad\rho(z, \bar z) \geq 0~.
}}
The normalization condition in the second line is a consequence of our discussion around~\ReggelimitS . 
We see that~\crossingequation\ is nicely satisfied for $\beta=1$, which is the self-dual point. We can now imagine expanding around this point to derive constraints on the possible distributions $\rho$ and the possible values of $k$.  

\newsec{Unitarity}

In this section we discuss two constraints from unitarity on the asymptotic distribution, $\rho$, of excess zeros. Both of the constraints  follow from the fact that the residue at any pole of the amplitude is a finite sum of Legendre polynomials with positive coefficients. Our first constraint concerns with the structure of the amplitude at large $s,t$. In particular, it has nontrivial implications for the rate of growth of the  leading trajectory $j(t)$.  The second is a constraint the region of support of the distribution $\rho(z,\bar z)$.  

To connect the discussion on the asymptotic distribution $\rho(t,z,\bar z)$ with unitarity we start from the product representation~\productreps. As $t\to m_i^2$ the amplitude factorizes as  
\eqn\identity{
\lim_{t\to m_i^2}A(s,t)={ c_i\over t-m_i^2} \prod_{n=1}^{L_i} \left(z^{e}_n(t=m_i^2) - s \right) = {1\over t-m_i^2}\sum_{j=0}^{L_i} f_{m_i^2 , j }^2 P_j\left(1 + {2 s \over m_i^2 - 4 m^2}\right)~,
}
where $\{z^e_l\}_{l=1}^{L_i}$ are the set of excess zeros at $t=m_i^2$. At large $t$, the set of excess zeros typically fluctuate a lot from one pole to the next. For example, the pole  $m_i^2$ may be on the leading trajectory where $L_i=j(m_i^2)$, while the next pole may belong on a sub-leading trajectory with $L_{i+1}\ll j(m_{i}^2)+1$. Hence, the distribution of the excess zeros at a specific pole does not have to coincide with the asymptotic distribution.\foot{As $t$ increases from $m_i^2$ to $m_{i+1}^2$, some excess zeros may move to infinity.} 

Instead, the asymptotic distribution is defined by coarse graining over the Legendre polynomials that appear in resonances spreading over scales that are larger than the separation between two poles on the leading trajectory. As discussed above, one way to preform that smearing is to analytically continue the amplitudes to $A\left(s(1+i\epsilon),t(1+i\epsilon)\right)$. Another way to coarse grain is to replace the amplitude by a different function $A(s,t)\to\widetilde A(s,t)$ for which the set of excess zeros does not fluctuate much from one pole to the next and at the same time leads to an identical asymptotic distribution. We can construct such a function $\widetilde A(s,t)$ from $A(s,t)$ by simply collapsing all the poles of the sub-leading trajectories to the nearest pole on the leading trajectory. More generally, we may collapse all the poles over some scale such that the corresponding excess zeros do not fluctuate from one pole of $\widetilde A$ to the next. Since the asymptotic distribution is a smooth function, it is not sensitive to the microscopic positions of the poles. Therefore, the distribution of the excess zeros of $\widetilde A$ would coincide with $\rho(t,z,\bar z)$. Of course $\widetilde A$ is only asymptotically crossing symmetric.

An important property that is preserved in this coarse graining procedure is that {\it the asymptotic distribution of excess zeros can be viewed as if it arises from a sum of Legendre polynomials with positive coefficients.}
Now we turn to studying some of the properties of sums of Legendre polynomials with positive coefficients.

\subsec{Positive Derivatives}

As a first step, let us note that we can always dimensionally reduce the problem to three dimensions. The positive sum of Legendre polynomials is then replaced by a positive sum of Chebyshev polynomials (this is because all the Clebsch-Gordan coefficients are non-negative) 
\eqn\reduce{
P_l\left(\cosh(\theta)\right)=\sum_{k=0}^{l} C^2_k \cosh(k\, \theta)\ ,\qquad C^2_k\ge0~. }
where the $\cosh(k\, \theta)$'s are the partial waves in three dimensions, also known as the Chebyshev polynomials, (viewed as functions of $\cosh\theta$). In the context of scattering amplitudes, at large $s,t$, we have $s = t \sinh^2 {\theta \over 2}$. 

Let us now consider a sum of partial waves with non-negative coefficients and take  the derivative of the logarithm of the sum   with respect to $\theta$. We get
\eqn\result{
\pa_{\theta} \log \left( \sum_{n=0}^{j(t)} C_n^2(t) \cosh n \theta \right) = { \sum_{n=0}^{j(t)} n\, C^2_n(t) \sinh n \theta \over  \sum_{n=0}^{j(t)} C_n^2(t) \cosh n \theta} >0\qquad{\rm for}\qquad \theta > 0.
}
It is useful to note a very simple upper bound that follows from replacing all the $\sinh$ factors in the numerators by $\cosh$ and by bounding the $n$'s in the numerator by the largest term $j(t)$. This gives
\eqn\resultabove{
\pa_{\theta} \log \left( \sum_{n=0}^{j(t)} C_n^2(t) \cosh n \theta \right) < j(t).
}

Similarly to~\result\ we get a positivity condition for the second derivative of the logarithm of the sum
\eqn\resultp{\eqalign{
&\pa_{\theta}^2 \log \left( \sum_{n=0}^{j(t)} C^2_n(t) \cosh n \theta \right) = { 1/2 \over \left(  \sum_{n=0}^{j(t)} C^2_n(s) \cosh n \theta \right)^2 } \times \cr
& \sum_{m,n=0}^{j(t)} 2n^2 C_n^4\delta_{n,m}+ (m^2\! +\! n^2)\,C^2_m C^2_n  \cosh (n\!-\!m) \theta + (m\,C^2_m\sinh m \theta - n\,C^2_n\sinh n \theta)^2 > 0 .
}}

These facts have interesting implications for the logarithm of the scattering amplitude.
Recall the representation 
$$\log A = F(t)+\int d^2z \rho(t,z,\bar z) \left[\log\left(z-s\right)-\log(z)\right]~.$$
Up to a multiplicative factor that depends only on $t$, this arises from an amplitude $A$ which can be decomposed as a sum over 
partial waves with positive coefficients. Therefore we can write 
\eqn\identsecf{\int d^2z\ \rho(t,z,\bar z) \log\left(z-s\right)=\log \left(\sum_{l=0}^{j(t)} C^2_l(t) \cosh(l\, \theta)\right)+\widetilde F(t)~.}
where $\widetilde F$ is some function of $t$ that will drop out after taking an $s$-derivative. 
It will be useful to record the transformation between derivatives with respect to $\theta$ and derivatives with respect to $s$. We have that $s = t \sinh^2 {\theta \over 2}$ and hence
\eqn\chain{\del_\theta=\sqrt{s(t+s)}\del_s~.}  

From~\result\ we see that if we take a derivative with respect to $s$ of the left hand side of~\identsecf, and use the first relation in~\chain\ then the result must be positive. Furthermore, using $\rho(t,z,\bar z)={j(t)\over t^2}\rho\left({z\over t},{\bar z\over t}\right)$ 
we find
\eqn\rewritecross{f(\beta)\equiv t^{1-k}\del_s \log A= \int d^2z {\rho(z,\bar z)\over \beta-z}\geq 0~.}

It is worthwhile to note a similarity with electrostatics. We can view $f(\beta)$ in~ 
\rewritecross\ as the electric field due to positive charges at the support of $\rho(z,\bar z)$. It obeys the equation $\bar \del f(\beta)=\rho(\beta,\bar\beta)$. We will soon see that this is a useful analogy to gain some intuition about the problem. Also, note that $\log A$, as written in~\distrcomp, is closely related to the potential due to the same charge distribution.

The upper bound~\resultabove\ implies 
\eqn\upperagain{\sqrt{\beta(1+\beta)}\int d^2z\ \rho(z,\bar z) {1\over \beta-z}<1~.}
In other words, combining~\rewritecross\ and~\upperagain\ we get $0\leq \sqrt{\beta(1+\beta)}f(\beta)<1$.

From~\resultp\ we obtain another interesting constraint on $f(\beta)$.  The bounded quantity is also monotonically increasing:
\eqn\conditionsDistr{\eqalign{
\pa_{\beta}\left( \sqrt{\beta(1+\beta)}f(\beta)\right)\geq 0\ .
}}

We can use~\rewritecross\ to expand $f(\beta)$ at large $\beta$. We find \eqn\moment{f(\beta)={1\over \beta}-{M_1\over \beta^2}+... ~.} 
with $M_1=-\int d^2z \rho(z,\bar z)z$ being the dipole moment of the distribution. Plugging this into~\conditionsDistr\ we obtain an inequality on the dipole moment $M_1$,
\eqn\dipolebound{M_1\geq {1\over 2}~.}
Therefore, the distribution has to have support mostly on the negative part of the real axis and it should certainly have some support for $z\leq -1/2$.

Let us now consider the limit of small $\beta$. In this limit we can use crossing symmetry (or equivalently~\crossingequation) to predict 
\eqn\fRegge{f(\beta)=-k\log(\beta)\beta^{k-1} + (k+1) M_1\beta^{k}+\dots\ .}
This is consistent with~\conditionsDistr\ only if \eqn\kbound{k> \half~.}
This is already a nontrivial constraint on the allowed form of  the Regge trajectory; we have to have $j(t)$ growing at least as fast as $\sqrt t$.

\subsec{The Support of the Distribution}

In this section we argue that the asymptotic distribution of excess zeros $\rho(z,\bar z)$ can only have support inside a certain ellipse extending between $z=-1$ and $z=0$, and
inside the unit disc. The argument again relies on the representation of the asymptotic amplitude in terms of a sum of Legendre polynomials with positive coefficients and its behaviour in the Regge limit \definitionRegge.

As we argued above, in the regime of large $s,t$, the scattering amplitude is controlled by the sum
\eqn\partialwave{
A(s,t)\simeq \widetilde A(s,t) \propto \int\limits_0^{j(t)} d j \  c_{j}(t) \ P_{j} \left(1 + {2 s \over t}\right)\ ,\qquad c_j(t)\ge0\ ,}
where we have replace the sum \identity\ at large $t$ by an integral. Here, the proportionality factor may depends on $t$. 

Let us understand how the Regge limit in both channels is reproduced. When $s \gg t$, the leading asymptotic simply comes from the Legendre polynomial of the highest order $j(t)$ and therefore we find $s^{j(t)}$ as expected. However, it is not guaranteed that this behavior persists in the asymptotic Regge limit. If we fix the ratio ${s \over t}\gg1$ then the upper limit of integration has a nontrivial scaling as we take $s \to \infty$. For the leading asymptotic to still come from the highest spin in the integral, $j=j(t)$, it is important that the coefficients $c_j(t)$ do not decay too fast. More precisely, our assumption number 6 in the list at the introduction translates into the following condition on the large $j$ behaviour of the coefficients
\eqn\decaybound{-\log c_j(t)<j\log(j(t)/j)\ ,}
at large $j$, smaller than $t^k$.

The opposite limit $t\gg s$ is less trivial. In this limit the argument of the Legendre polynomials is very close to one and we can approximate \partialwave\ by\foot{A simple way to derive this is to use the integral representation of the Legendre polynomyals $P_j(1+\epsilon)=\int_0^1dx[(1+\epsilon)-\sqrt{(1+\epsilon)^2-1}\cos(\pi x)]^j\simeq \int_0^1dx\,e^{\sqrt{2\epsilon}j\cos(\pi x)}=I_0(j\sqrt{2\epsilon})$.}
\eqn\partialwaveB{
A(s,t) \simeq \int\limits_0^{j(t)}\! d j \  c_{j}(t)\, I_{0}(2 j \sqrt{s/ t}).
}
where $I_0(x)$ is the Bessel function. 

If instead, we start from the dimensionally reduced amplitude to three dimensions,~\partialwave\ is replaced by 
\eqn\partialwaveC{A(s,t)\simeq\int\limits_0^{j(t) }dj \ c_j(t)\cosh\left[j\cosh^{-1}\left(1+{2s\over t}\right)\right]\ .} 
Similarly, in the limit of $t\gg s$ it is approximated by 
$$A(s,t)\simeq\int\limits_0^{j(t) }dj \ c_j(t)\cosh\left(2j\sqrt{s\over t}\right)~.$$ 
In either case, at large positive $j$ we can further approximate the integral by\foot{The same asymptotic structure holds true in four dimensions, $\log I_{0}\left(2 j \sqrt{ {s \over t} }\right) \sim 2 j \sqrt{ {s \over t} }$.}
$$A(s,t)\simeq\int\limits^{j(t) }dj \ c_j(t)\,e^{2j\sqrt{s\over t}}~.$$
We conclude that to reproduce the Regge behaviour $A\simeq t^{j(s)}$ in the limit $t\gg s$ we have to require that
\eqn\tchannelregge{\int\limits^{j(t) }dj \ c_j(t)\,e^{2j\sqrt{s\over t}}\ \sim\  t^{j(s)}~.}
This equation is true in any number of dimensions.

In order to achieve~\tchannelregge, the integral should be dominated at large $j$, where we can evaluate it by a saddle point. We will now focus on $j(t)= t^k$ to simplify our considerations.
The correct behaviour is reproduced if the location of the saddle point  scales like  $j_*\propto s^{k - {1 \over 2}} \sqrt{t} \log t$.
Note that this saddle point is within the region of integration only if the leading trajectory grows at least as fast as \eqn\anotherbound{j(t)>\sqrt{t} \log(t)~,}
which is indeed the case for $k>1/2$ \kbound.  
We need to require that $\log c_j(t)$ is such that the integral is stationary at the saddle point and thus we find 
\eqn\sadpoint{{\del\over\del j }\log(c_j(t))\biggr|_{j_*}+2\sqrt{s/t}= 0~,}
and, in addition, \eqn\sadvalue{\log c_{j_*}(t)+2j_*\sqrt{s/t}=j(s)\log(t)~.}

Since these equations depend on the ``auxiliary'' variable $s$, they suffice to fix the form of $c_j(t)$ at large $j$ and $t$. From~\sadpoint\ we find
$$\log c_j(t)\simeq{1-2k\over 2k}j^{2k\over 2k-1}t^{k\over 1-2k}\left(\lambda\log(t)\right)^{1\over 1-2k}~,$$
where $\lambda$ is a constant. From~\sadvalue\ we find after a short computation that
$\lambda=k$ and thus\foot{
 Note that since the omitted constant under the logarithm in $j_*\propto s^{k - {1 \over 2}} \sqrt{t} \log t$ could be $\log(t/s)$ 
and ~\tchannelregge\ is only valid for $t\gg s$, we can obtain $c_j(t)$ only for a range of $j$'s much smaller than $j(t)$. }
\eqn\generic{\log  c_j(t)\simeq{1-2k\over 2k}j^{2k\over 2k-1}t^{k\over 1-2k}\left(k\log(t)\right)^{1\over 1-2k}~.}
For example, setting $k=1$ in \generic, we correctly reproduce the result for partial waves of the Veneziano amplitude~\SiversIG. The estimate \generic\ is only the 
leading piece in the logarithm of $c_j(t)$ for large $s,t$. There may well be subleading corrections in $s,t$ and also corrections that are not exponentially large and hence do not influence the saddle point equations.

The coefficients $c_j(t)$ can be interpreted as the  amplitude for the decay of a resonance of spin $j$ and mass $\sqrt t$ to our external particles.  (This can be seen by setting $s=0$ in~\partialwave\ and considering the optical theorem in the t-channel.) We see that it exponentially decreases with $j$ (this follows for example from~\anotherbound\ and the fact that $k>1/2$).

We are now in position to study the distributions of zeros generated by~\partialwaveB\ or~\partialwaveC. The support of such distributions  depends on the properties of $c_j(t)$, but there is also some degree of universality.
Many ``generic'' choices for the coefficients lead to the same result, and we will in particular discuss the choice that is dictated by having Regge physics,~\generic.

Let us start from some examples.

\medskip 

\item{A.} A single Legendre polynomial, $P_j\left(1+{2s\over t}\right)$ has all its zeros between $-1<{s\over t}<0$ and thus the distribution $\rho(z,\bar z)$ would have its support between $z=-1$ and $z=0$. All the zeros are on the real axis. 

\item{B.} If we assume that the coefficients decay very fast, e.g. $c_j(t) = {1 \over j!}$, then we get a distribution that is supported along a complicated curve in the complex plane. Such examples however do not  behave as in \generic\ or \decaybound\ and, hence, fail to Reggeize.

\ifig\figdistas{Distribution of zeros for \generic\ for $k = {3 \over 4}$ and the maximal spin being $400$. } {\epsfxsize 3.3in \epsfbox{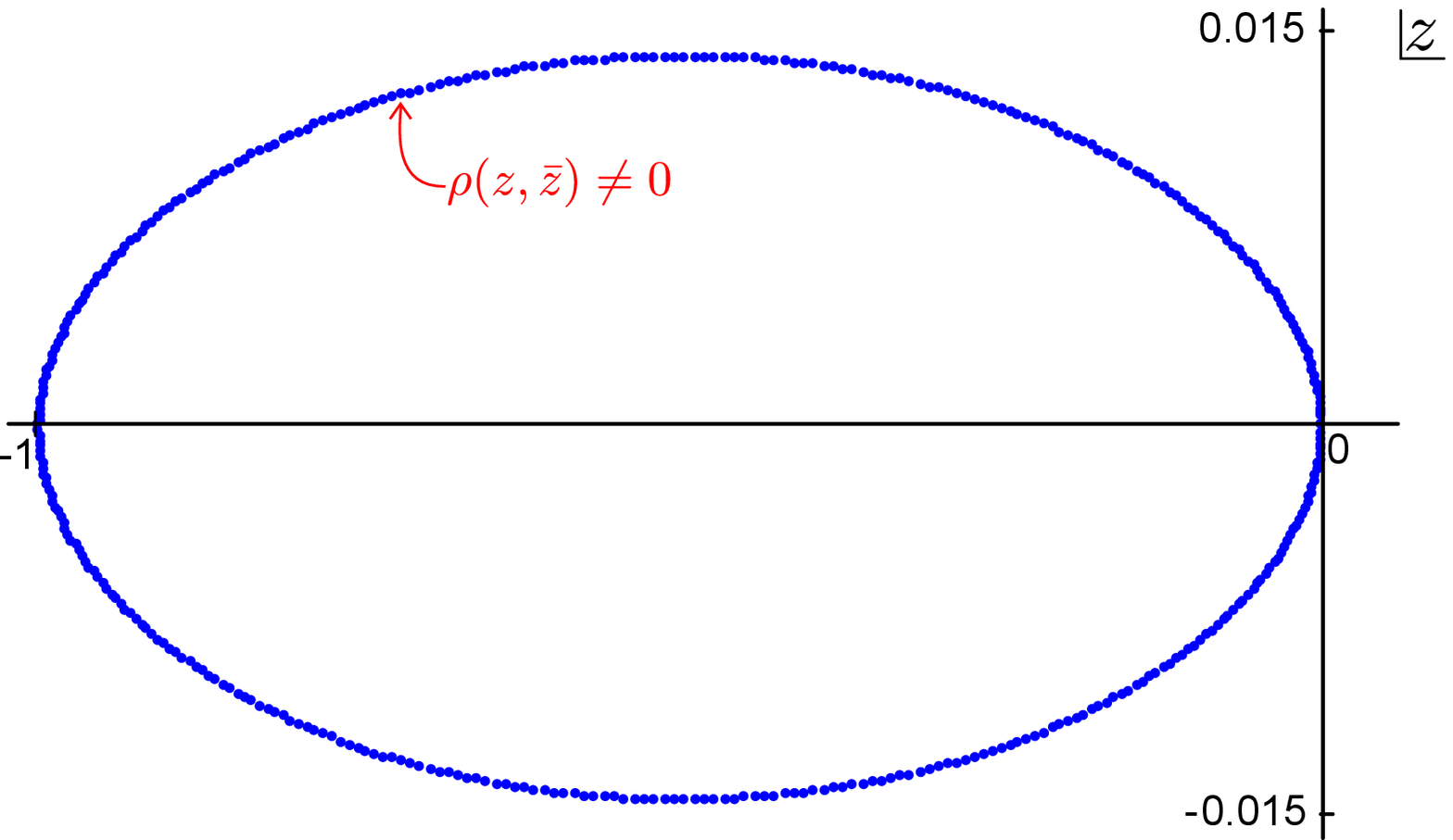}}

\ifig\figdistasnoise{Distribution of zeros for \generic\ for $k = {3 \over 4}$ and the maximal spin being $400$ with the random noise which is $10 \% $ of \generic . } {\epsfxsize 3.3in \epsfbox{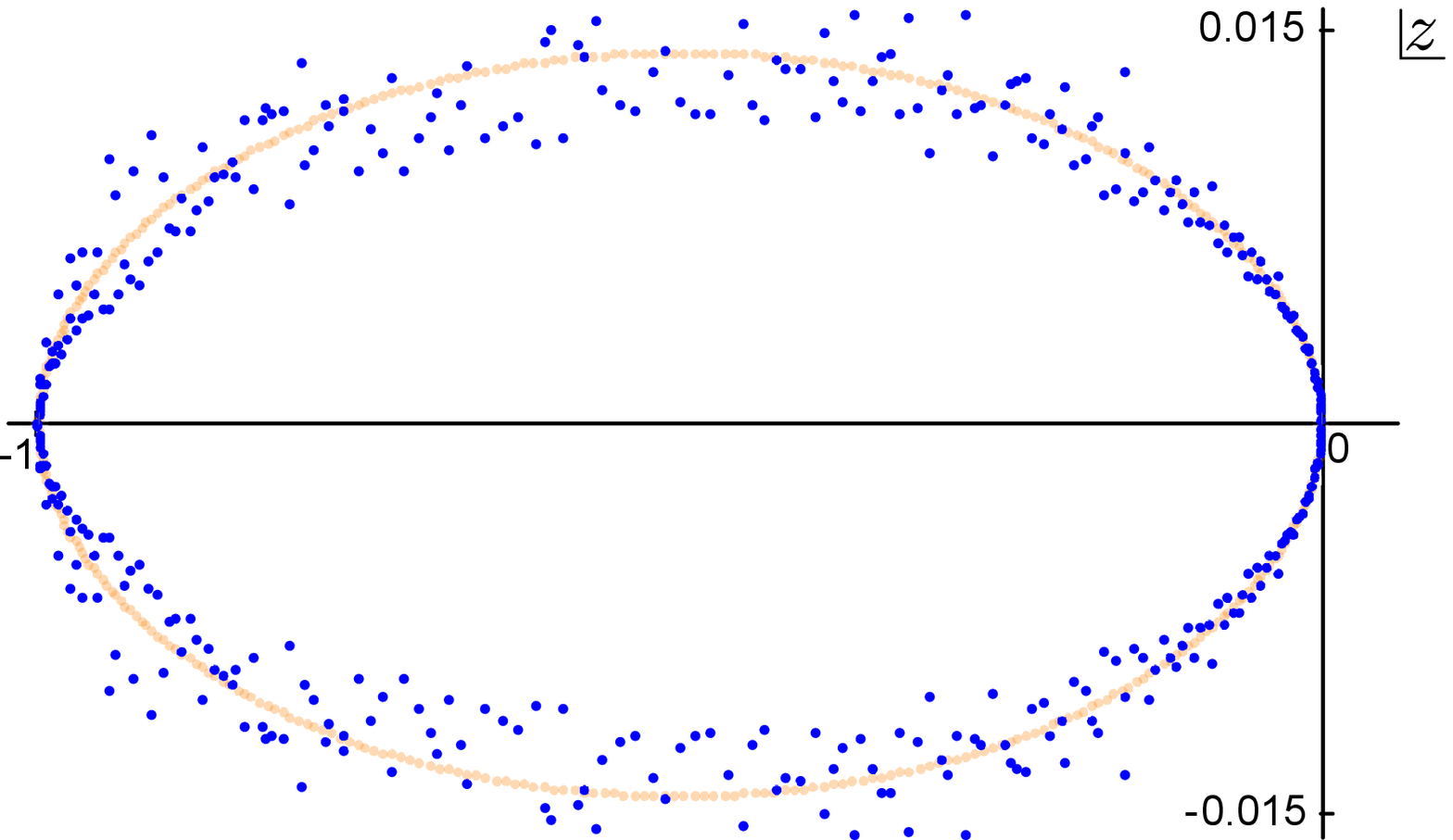}}

\item{C.} Generic choices of the coefficients $c_j(t)$, and in particular ones that behave as in~\generic, generate smooth distribution of zeros
{\it localized within an ellipse between $z=-1$ and $z=0$ and inside the unit circle}, (see figure \figdistas\ and \figdistasnoise\ for one such example). This fact will be of utmost importance to us. It is easily verified with some numerical experimentation but we leave a detailed rigorous proof to the future. 

\medskip

 \noindent Closely related problems of the distribution of zeros of random polynomials have been extensively studied in the mathematical literature, see for example \refs{\zeros,\zerosort} and references therein. Universality phenomenon in the distribution of zeros that we observed above is well-known as well, see e.g. \Tao. In this language, we assume that the sum of Legendre polynomials that produces the asymptotic amplitude is in the universality class of examples A and C above (or, more generally, in the class where zeros condensate inside the unit disc $|z|<1$ and possibly $z=-1$), which is motivated by the Regge limits and numerical experimentation. It would be desirable to put this on a more solid mathematical ground.

\subsec{Non-zero Support Outside of the Unit Circle}

It is very easy to generate an amplitude that satisfies the constraints of unitarity if we relax the condition that the density of zeros is localized within the unit circle. As a simple example consider the following model 
\eqn\simplemodeldistr{
\pa_s \pa_t \log A =k {\alpha^k \over 1 + \alpha^k} s^{k-1} t^{k-1} \left( {1 \over (s+ \alpha t)^k} + {1 \over (t + \alpha s)^k} \right) ,
} 
where without loss of generality we can assume that $\alpha > 1$.

Using \simplemodeldistr\ we can compute the density of the excess zeros as explained above. It has
non-zero support along the real axis starting from $0$ and until $- \alpha$. One can easily check
that this amplitude satisfies all the unitarity constraints for some range of $\alpha$ and $k \neq 1$
(for example $\alpha=2.8$, $k=0.9$). However, the corresponding distribution of zeros does not
arise from sums over Legendre polynomials of the type that appear in our problem. In our
problem, the distributions are localized into the half-disk. If the analytic continuation of
the asymptotic limit of the amplitude around the positive axis leads to a function with
branch cuts away from the half-disk, then it cannot be re-constructed (even near the
positive axis) from a distribution in the unit disk.

\subsec{An Argument for $k\leq 1$}

We are concerned with the equation $\bar \del f(z)=\rho(z,\bar z)$ corresponding to a positive distributions of charges which create some electric field. This equation is solved by (up to anti-holomorphic functions that we do not need to keep track of)~\rewritecross.

As we have just seen, $\rho(z,\bar z)$ only has support for ${\rm Re}(z)\leq 0$ and, by construction, $\rho$ is real and positive and symmetric about the real axis. Finally, the distribution is normalized $\int d^2 z\, \rho(z,\bar z)=1$.

We start from some very simple comments. Take $z=\beta>0$ on the real axis and consider
$$f(\beta)=\int\limits_{Re(z')\leq 0}\!\! d^2z'\, {\rho(z',\bar z')\over \beta-z'}~.$$ 
Because of the symmetry of $\rho$ we see that ${\rm Im} f(\beta)=0$. We write the real part
\eqn\deno{ Re f(\beta)=\int\limits_{Re(z')\leq 0}\!\! d^2z'\, \rho(z',\bar z'){\beta-Re(z')\over |\beta-z'|^2}>0\ .}
This is clearly positive because $\beta$ is positive, $\rho$ is positive, and $-{\rm Re}(z')$ is nonnegative.  
Indeed, as we have argued in subsection 4.2 the distributions of the type which arise in our problem have support in the unit half-disc.
So one can immediately establish 
\eqn\upper{k\leq 1}
as long as the distribution of zeros has some support away from the imaginary axis, which is a consequence of \dipolebound. This conclusion follows from the Regge limit~\fRegge, which shows that for $k>1$ the function $f$ goes to zero at $\beta=0$. However, the  presence of charges at negative ${\rm Re}(z')$ and their absence at positive ${\rm Re}(z')$ cannot be consistent with a vanishing electric field at $\beta=0$.

\newsec{An Argument for the Linearity of Regge Trajectories}
 
Until now we used unitarity of the amplitude \identity\ and crossing symmetry in the Regge limit \generic\ to constrain the region of support of the asymptotic distribution $\rho(z,\bar z)$ \distrcomp. We have given some arguments for why $\rho(z,\bar z)$ only has support inside an ellipse with ${\rm Re}(z) \leq 0$ and moreover touches the unit disc only at $z=-1$. We shall now combine these ideas with analyticity of the amplitude in the large $s,t$ asymptotic region to rule out a non-linear leading trajectory.
 
 Our analysis so far has led to~\kbound,\upper 
 \eqn\soweit{\half< k\leq 1} and we now complete the analysis by showing that the only consistent choice is $k=1$.
  
Let us first discuss the analytic properties of the function $f(\beta)$, \rewritecross. It is related to the amplitude as $f(\beta)=t^{1-k}\partial_s\log A$, where this relation holds in the $|s|,|t|\gg1$ asymptotic regime in a small wedge around the real axis (${\rm Arg}(s),{\rm Arg}(t)>\epsilon$, see figure 3). After we analytically continue $f$ outside that wedge, there is no reason for it to agree with the amplitude. In other words, the analytic continuation does not commute with the asymptotic limit where $f(\beta)$ is defined. The fact that analytic continuations and asymptotic limits do not commute is a standard phenomenon. 
In our context, this happens because at large positive $s,t$ the function  $f(\beta)$ is dominated by the excess zeros, while for negative $t$ this is  not the case. 

The function $f(\beta)$ can be analytically continued using the representation~\rewritecross,
$$f(\beta)=\int d^2z{\rho(z,\bar z)\over \beta-z}~.$$ 
As long as we stay away from the region where $\rho(z, \bar z)$ has support, $f(\beta)$ has no singularities. Inside the unit half disc, where $\rho$ has support, it has cuts at positions and strength that is dictated by $\rho(z,\bar z)$. To understand better how this works, we first construct from $f(\beta)$ a new function with simple crossing transformation in the $s,t\gg 1$ asymptotic regime\foot{Note that crossing $s\leftrightarrow t$ maps the asymptotic regime of $s,t\gg1$ back to itself. Hence, this symmetry of the amplitude is respected by the analytic continuation of the function $h(\beta)$.}
\eqn\hfunction{
h(\beta)\equiv-[\beta\partial_\beta+(1-k)]f(\beta)=t^{2-k}\partial_s\partial_t\log A(s,t)~.
}
$h(\beta)$ inherits its analytic properties from $f(\beta)$. In particular, $h(\beta)$ is analytic and single-valued outside of the region of non-zero support of the distribution. The transformation of $h(\beta)$ under the exchange of $s\leftrightarrow t$ is quite simple 
\eqn\hcrossing{h(1/\beta)=s^{2-k}\partial_s\partial_t\log A(s,t)=\beta^{2-k}h(\beta)~.}

One immediate implication of \hcrossing\ is that $h$ has a branch point of degree $k$ starting at $\beta=0$. This cut can only end at $\beta=-1$. It is because the crossing transformation \hcrossing\ relates the analytic continuation of $h$ inside the circle to its value outside, where it is analytic. Moreover, the region of support of the distribution $\rho$ only touches the unit circle at -1. Hence, we can write $h(\beta)$ as
\eqn\pfunction{h(\beta)=k\left({\beta\over1+\beta}\right)^k{g(\beta)\over\beta}~.}
where $g(\beta)$ is invariant under crossing $g(\beta) = g({1 \over \beta})$ and is analytic outside of the region of  support of the distribution. From \fRegge\ and \hcrossing\ we learn that $g(0)=g(\infty)=1$.   Furthermore, crossing transfers the nice analyticity properties of $g(\beta)$ from large $\beta$ to small $\beta$.

We notice that analyticity and crossing together are sufficient to fix $g(\beta)$. The argument for that goes as follows. Let us define a new function $\tilde g(\beta)$. Outside of the region with non-zero support of the distribution we have $\tilde g(\beta) = g(\beta)$ and inside the distribution support region we define
\eqn\definitiong{
\tilde g(\beta) = g \left({1 \over \beta} \right) .
}

Defined in this way, $\tilde g(\beta)$ is analytic in the whole complex plane except for the crossing symmetric point $\beta=-1$, where it may be singular. For example, if the density of zeros has a delta function at $z=-1$, 
$\rho\sim\partial^{n}\delta^{(2)}(z+1)$, then this would lead to a pole at $\beta=-1$ in $g(\beta)$.
Such singularities however cannot come from the positive sums of Legendre polynomials of the type discussed in the previous section.\foot{Moreover, if instead of external flavoured scalars, we would consider the $2\to2$ scattering of identical particles, then all odd spins would automatically decouple. In such case, the distribution of excess zeros near $z=-1$ would be identical to the behaviour near $z=0$ and thus $h(\beta)$ would just have a branch point there.}
 
Thus, we conclude that $\tilde g(\beta)$ is analytic in the whole complex plane and thus $\tilde g(\beta) = 1$. Hence, also $g(\beta)=1$ outside of the region where the distribution has support. 

Notice that it is absolutely crucial for the argument that the original function is analytic at $|\beta|=1$, $\beta\ne-1$. As explained in the previous section, this is a consequence of unitarity. It is the reason we can analytically continue $h$ across $|\beta|=1$ without hitting a singularity.

Using~\pfunction\ and~\hfunction\ we can infer  $f(\beta)$. It is given by solving the equation 
\eqn\diffeq{\left[-\beta\del_\beta+(k-1) \right]f(\beta)={k\over \beta}\left({\beta\over 1+\beta}\right)^k\ ,\qquad\lim_{\beta\to\infty}\beta f(\beta)=1~. }
This can be easily integrated to give 
\eqn\toymodelfbeta{f_k(\beta) ={1 \over \beta} \ _2 F_1 (k,k,1+k, - {1 \over \beta})\ .}
Let us reiterate that \toymodelfbeta\ is valid everywhere outside of the region where the distribution has support. However, only for $\beta$ on the positive real axis this describes the asymptotic form of the original scattering amplitude.

We will now argue that~\toymodelfbeta\ is only consistent with unitarity and positivity of the distribution if $k=1$. Expanding~\toymodelfbeta\ at small beta, we have
\eqn\unitarityconstraint{
f_k(\beta)={1\over \beta}-{k^2\over k+1}{1\over \beta^2}+...~.
}
Comparing this with our constraint~\dipolebound, we infer that $k\geq 1$. Together with~\soweit\ we thus conclude that $k=1$, as promised.

For the sake of completeness
we quote here the density of zeros that correspond to~\toymodelfbeta 
\eqn\distr{\rho_k(x)={k \sin(k \pi) \ (-x)^{k-1} \over \pi} \left(- \log(- x) + \sum_{m=1}^{\infty} {(k)_m (1 - (-x)^m ) \over m \ m!} \right)~ .}
The moments of this distribution are $M_n ={k \over k+n} {\Gamma(k+n) \over \Gamma(k) \Gamma(1+n)}$.
These are only consistent with the zeros of positive sums of Legendre polynomials (with the other constraints we discussed) if $k= 1$.

In the case $k=1$~\toymodelfbeta\ simplifies  to 
\eqn\venef{f_{k=1}(\beta)=\log\left({1+\beta\over \beta}\right)~.}
 Accordingly, the distribution $\rho$ in~\distr\ simplifies. One has to take the limit $k\rightarrow 1$ in~\distr\ carefully since the prefactor vanishes in this limit. One finds 
\eqn\densvene{\rho_{k=1}(x)=1~.}

The results~\venef\ and \densvene\ allow us to fix the amplitude for large positive $s,t$. For this we recall~\rewritecross, which gives  $\del_s \log A = \log\left({s+t\over s}\right)$. This is solved by 
$$\log A=(t+s)\log(s+t) -s\log s+f(t)$$
and the function $f(t)$ is now fixed uniquely by crossing to be $f(t)=-t\log t$ and thus we finally find
\eqn\gm{\log A=(t+s)\log(s+t) -s\log s-t\log t~.}
This is precisely what we would have got from the Veneziano amplitude in the limit of large positive $s,t$ (see appendix C).\foot{Actually, in the Veneziano amplitude this is also valid for negative large $s$ or $t$, where this coincides with the hard-scattering or the Gross-Mende regime~\GrossKZA, \GrossGE. In general, as we have explained, we do not expect this  formula to be a valid asymptotic estimate for negative $s$ or $t$ but only when both are large and positive.} There is therefore only one consistent asymptotic form for the amplitude at large $s,t$, namely,~\gm.

To summarize, we make some additional comments on the results in this section. $k=1$ implies that the leading Regge trajectory is asymptotically linear and thus the fastest spinning resonances are asymptotically equidistant. But the most general result that we have shown is that any consistent amplitude satisfying our assumptions takes the form~\gm\ for large positive $s,t$. This implies that the trajectories are asymptotically linear but it actually leads to stronger consequences. One consequence was discussed around~\decayingamp\ where we have shown that~\gm\ leads to the existence of strings in the theory. Let us now note also that~\gm\ cannot be reproduced from a single Regge trajectory. However we can use the estimate~\generic\ to conclude that~\gm\ implies the existence of infinitely many asymptotically parallel Regge trajectories (i.e. there must be infinitely many asymptotically linear, parallel daughter trajectories). 

\newsec{Conclusions}
\subsec{Summary}
In this paper we considered scattering processes of weakly coupled particles that involve the exchange of massive higher-spin resonances. Such scattering processes are described by meromorphic, crossing symmetric scattering amplitudes with the residues being  sums of Legendre polynomials with non-negative coefficients. Imposing that such an amplitude does not grow too fast at high energies makes the problem very constrained. For example, an immediate consequence  is that there is an infinite number of particles with arbitrarily high spin in the spectrum.

There are several known solutions to that problem. All of them are theories of strings: either fundamental tree-level string theory or strings of Large-N confining gauge theories. The corresponding scattering amplitudes are generically sensitive to the details of the underlying theory. On the other hand, in all known cases the asymptotic form of the leading Regge trajectory $j(t)$ for large $t$ is believed to be linear. This regime corresponds to $s$ and $t$ being large and positive. In this work we have sown that unitarity and crossing constraints are sufficient to fix the amplitude uniquely in this regime. We have found that the only unitary crossing symmetric solution in this asymptotic regime coincides with the limit of the Veneziano amplitude \uniqueampl . In particular, this should be true in pure Yang-Mills theory at large $N$. Thus, to leading order,  the Regge trajectory $j(t) =\alpha' t + o(t)$ is linear. It  implies that in some sense all theories of weakly coupled higher-spin particles are theories of strings.

\subsec{Future Directions}

A promising future direction is to study corrections to the leading universal asymptotic behaviour of the amplitude. Is there some universality in the leading corrections as well?  Even if some of the corrections are universal, it is clear that beyond some point the amplitude will not be universal. To fix the amplitude beyond this point we must add some additional input / assumptions. 

An analogous situation occurs in the context of the conformal bootstrap (see~\RychkovIQZ\ for a review). There, to strengthen the results, one often assumes that there is a gap in the spectrum of operators. From unitarity and crossing one then produces constraints on the dimensions and three-point coefficient of heavier operators. It is natural to ask: what would be the appropriate physical input to add to the amplitudes bootstrap? Instead of dimensions of operators we have masses of particles in the context of the scattering amplitudes. Our analysis, however, is asymptotic in nature and is not sensitive to the spectrum of light particles. So an interesting direction would be to understand if one can extract any mileage from the bootstrap for the physics of light particles. Progress along these lines was recently achieved in~\refs{\JoaoPedro}. 

It could also be that one can understand the subleading corrections to $j(t)$ by adding information about the low lying particles in the theory. This happens in the conformal bootstrap: there is an interplay between very large spin operators and low spin operators~\refs{\FitzpatrickYX,\KomargodskiEK}.

Many of the  CFTs are isolated (sitting at corners of the bootstrap bounds) while amplitudes are not (for example, we can add some very massive fields to pure Yang-Mills theory).
Are there some conditions that would select, say, pure Yang-Mills similarly to the case of the 3D Ising model~\ElShowkHT?
One possibility might be to constrain the value $a_{\rm UV}$ of the $a$-anomaly at short distances.
Finding how this coefficient is encoded in the S-matrix of mesons would be an important step forward.

We also have to understand what are the effects of the degeneracies in the spectrum. In the case of tree-level string theory we have an exact, large, degeneracy in the spectrum. However, it is likely that generically, say in Yang-Mills theory, all degeneracies are lifted. Can we use unitarity and crossing to restrict the patterns of degeneracy lifting? This seems to only enter at the $O(1)$ contribution to the trajectory $j(t)$.

Of course, there are many other extensions one can consider. For example one can study unitarity and crossing constraints for $2\to2$ scattering amplitudes of particles with spin (here we only considered constraints coming from scalars amplitudes). One can also consider mixed bootstrap  for the combinations of scattering amplitudes that form closed sub-sectors similar to~\KosBKA. It would be also interesting to incorporate the constraints due to higher-point scattering.

Another expectation is that having  a massless graviton in the spectrum should be much more constraining. One can imagine that classical string theory is the unique theory of massive, stable, higher-spin particles that contains gravity. 
See~\NimaYutin\ for recent work on the subejct.

It would be also interesting to understand if our universal asymptotic formula implies something interesting about CFTs. Indeed, the flat space scattering amplitudes are related to the Mellin amplitudes in a particular limit \PenedonesUE. Another direction is to understand if our results have any implications for cosmology \ArkaniHamedBZA .

More broadly, it would be interesting to identify a set of  questions for which general principles like unitarity and crossing are useful and constraining. In the context of the $S$-matrix bootstrap, despite the long history of the subject, understanding the implications of the constraints due to unitarity, crossing, and analyticity  is still largely an open problem.\foot{See, for example, question number 72 by Juan Maldacena in \StromingerTalk\ and appendix~G in \MaldacenaSF.} 

\newsec{Acknowledgments}

We thank Ofer Aharony, Nima Arkani-Hamed, Matthijs Hogervorst, Yu-tin Huang,  Baur Mukhametzhanov, Jacob Sonnenschein and Shimon
Yankielowicz for useful discussions.
SCH's research was partly funded by the Danish National Research Foundation (DNRF91).
Z.K. is supported in part by an Israel Science Foundation center for
excellence grant and by the I-CORE program of the Planning and Budgeting Committee
and the Israel Science Foundation (grant number 1937/12). Z.K. is also supported by the
ERC STG grant 335182 and by the United States-Israel BSF grant 2010/629. A.S. has been supported by the I-CORE Program of the Planning and Budgeting Committee, The Israel Science Foundation (grant No. 1937/12) and the EU-FP7 Marie Curie, CIG fellowship.

\appendix{A}{Scattering Amplitudes Primer}

\subsec{Resonance Exchange}

Let us review some of the basic properties of the $2\to2$ scattering amplitude where all the external particles are taken to be scalar particles. The unique interaction vertex containing twice the scalar particle $S$ and a spin $L$ particle, $\phi_{\mu_1,...,\mu_L}$, in the symmetric traceless representation\foot{Here we consider a complex scalar $S$ instead of a real scalar, so that it could couple to both even and odd spins.} is
\eqn\vertex{f_{SS\phi}(\del^{\mu_1}\cdots\del^{\mu_L}  S)S^\dagger   \phi_{\mu_1...\mu_L}~.}
All the other interaction vertices can be simplified to this form using the equations of motion and the Fierz-Pauli transversality condition
$$\square\, S=m^2_{S} S\ ,\qquad \square\, \phi_{\mu_1...\mu_L}=m^2_\phi \phi_{\mu_1...\mu_L}\ ,\qquad \del^{\mu_1}\phi_{\mu_1...\mu_L}=0\ .$$
That means that the contribution of other vertices to the S-matrix would differ, at most, by a polynomial in momentum.

Exchanging the resonance $\phi$ in the s-channel and using the vertex~\vertex\ we get the contribution to the scattering amplitude
\eqn\onere{
A(k_i)\sim f^2_{SS\phi}\left[\prod_{i=1}^L(k_1-k_2)^{\mu_i}(k_3-k_4)^{\nu_i}\right]\times
\langle \phi_{\mu_1...\mu_L}(k_1+k_2)\,\phi_{\nu_1...\nu_L}(k_3+k_4)\rangle~. 
}
(To get to the form~\onere\ we used momentum conservation and the fact that the propagator is transverse.) Equation~\onere\  can be written using the usual Mandelstam variables $s=(k_1+k_2)^2$, $t=(k_1+k_3)^2$. This is easily done by first going to the center-of-mass frame where we can take $k_{1,2}=(E,0,0,\pm p)$ and $k_{3,4}=(-E, \pm p\hat n)$ with $\hat n$ a unit three-vector such that 
$\hat n\cdot \hat{z}=\cos\theta$. We see that the dependence on $\hat n$ in~\onere\ comes from the prefactor $(k_3-k_4)^{\nu_1}\cdots (k_3-k_4)^{\nu_s}$. This factor can be taken to be traceless due to the tracelessness of the propagator,  $$(k_3-k_4)^{\nu_1}\cdots (k_3-k_4)^{\nu_L}\quad\longrightarrow\quad  (k_3-k_4)^{\nu_1}\cdots (k_3-k_4)^{\nu_L}-{\rm traces}~.$$ Furthermore, the vector $k_3-k_4$ has only space-like components. Therefore, the dependence on $\hat n$ comes from a traceless symmetric tensor of spin $L$  contracted with a unit vector. This is precisely how Legendre polynomials arise\foot{In general space-time dimension $D$ this leads to the Gegenbauer polynomials, $$P_L(x)\longrightarrow\ \  _2F_1\left(-L,L+D-3, {D-2\over 2},{1-x\over 2}\right)~.$$
Legendre polynomials are recovered for $D=4$.} and therefore the scattering amplitude in the center of mass frame is 
\eqn\onere{A(k_i)\sim f^2_{SS\phi}{P_L(\cos\theta)\over 4E^2-m_\phi^2}~. }
Therefore, near the s-channel pole, the relativistic-invariant form of the scattering amplitude due to the exchange of a spin $L$ particle is 
\eqn\onerei{
A(s,t)\biggr|_{s\approx m_{\phi}^2} \simeq f^2_{SS\phi}{P_L\left(1+{2t\over m_\phi^2-4m_S^2}\right)\over s-m_\phi^2}~ \ .
}

\subsec{Resonances at $j(t) = n$}

Here we argue that at the points $\{t_n\}$ where $j(t_n)=n$ is a non-negative integer, $t$ hits a resonance. We start from the large $s$ Regge form of the amplitude \definitionRegge\ which we repeat here for convenience
\eqn\definitionReggetwo{
\lim_{|s| \gg |t|,m^2_S} A(s,t) = F(t) (-s)^{j(t)}\ ,\qquad{\rm arg}[s] \neq 0 \ .}

The function $F(t)$ must have poles when $t$ hits the mass squared of a resonance on the leading trajectory. Therefore, we can write \eqn\functionF{F(t)= {f(t)\over \sin(\pi j(t))}} where $f(t)$ is a regular function at $t_n$. In fact, let us consider the discontinuity in $s$ in~\definitionRegge. Using the fact that ${\rm Im}_s[ (-s)^{j(t)} ] = |s|^{j(t)}\sin(\pi j(t))$, we get that $F(t)|s|^{j(t)}\sin\left(\pi j(t)\right)=f(t)|s|^{j(t)}$. 
This has to be non-negative because from~\onereii\ we see that for positive $t$ and sufficiently large $s$ the imaginary part is always non-negative. Therefore, 
\eqn\positivepref{f(t)\ge 0~.}
Let us analyze carefully what happens near some $t_n$. Let us assume that $j'(t_n)>0$ (we will soon justify this by unitarity). Then for $t<t_n$ (sufficiently close to $t_n$) 
$ s^{-n}A(s,t)$ tends to zero as $|s|\rightarrow\infty$.  We can therefore use a dispersion relation and get for $t\rightarrow t_n^-$ that the contribution to the dispersion relation from large $s'$ is
$$s^{-n}A(s,t)\sim \sin(\pi j(t))F(t) \int ds' {s'^{j(t)-n}\over s'-s}$$ 
where we have used the fact that ${\rm Im}[ (-s)^{j(t)} ] = |s|^{j(t)}\sin(\pi j(t))$. This diverges logarithmically as we hit $t=t_n$. The integral from the large $s'$ region
gives a simple pole 
\eqn\pole{{f(t_n)\over j(t)-n}~,} which corresponds to a resonance at $t=t_n$ as predicted. It has spin $n$ since to obtain the amplitude we need to
multiply by $s^n$. The imaginary part in $t$ near this pole is proportional to $j'(t_n)$ and hence from unitarity
$$j'(t_n)>0\ ,\qquad {\rm for}\quad f(t_n)\ne0~.$$ Here we have used~\positivepref, which guarantees that the numerator in~\pole\ is positive.

\subsec{Impact Parameter Transform}

In this section we review the transform to the impact parameter amplitude. We argue that if the leading trajectory is linear then the inelastic part of the large energy, large impact parameter amplitude is dominated by \decayingamp.

The transformation between the momentum amplitude $A(s,t)$ and the fixed impact parameter amplitude is 
\eqn\impactp{
A(\vec b,s)= \int d^{2}\vec k_{\perp}e^{i\vec{k}_\perp\cdot \vec{b}} A(t,s)
}
where at high energy $t \simeq - \vec k_{\perp}^2$. We can rewrite~\impactp\ using the azimuthal symmetry
\eqn\impactpaz{A(b,s)\sim \int|k_\perp|d|k_\perp| J_0(|k_\perp|b) A(t,s)~.}
The integral in the $t$ variable is taken in the interval $[-s+4M_S^2,0]$. The contribution from $t\ll s$ takes the form 
\eqn\impactpi{A(b,s)\sim \int dt F(t) J_0(\sqrt{-t}b) e^{j(t)\log (-s)}}
There is a saddle point off the integration contour at 
$${\del\over\del t}\left(i\sqrt {-t} b +j(t) \log(-s)\right)=0~,$$
which gives ${ib\over 2\log(-s)}=\sqrt{-t}j'(t)$. The saddle point is therefore at imaginary $\sqrt{-t}$, which means positive $t$. This is the region where we expect the amplitude to be universal.
Suppose that at large $t$ we have $j(t)\sim\alpha't$ then we find that at the saddle point $\sqrt{-t}={ib\over 2\alpha'\log(-s)}$. This is self consistent if the 
impact parameter $b\gg\log s$ in units of the QCD scale. 
Plugging this saddle point back into the integral we find 
\eqn\decayingamp{
A(b,s)\sim e^{-{b^2\over 4\alpha' \log(s)}}~.
}
Notice that the real part of the amplitude receives the leading contribution from the pole closest to the real axis at large impact parameter, which generate the Yukawa potential of the form $e^{- m_{min} b}$. Indeed, we have that $m_{min} b \ll {b^2\over 4\alpha' \log(s)} \sim \sqrt{t} b$. On the other hand, the imaginary part is captured by \decayingamp\ since the contribution from the poles that we get when shifting the contour is purely elastic. For a detailed discussion of this in the case of string theory see e.g. \AmatiWQ .

\subsec{Mandelstam Argument}

One notable development in Regge theory is the so-called ``Mandelstam Argument''~\mandelstam. Here, we review the argument and its possible loopholes.

One assumes that
$j(t)$ is analytic in the complex $t$-plane and does not grow faster than $j(t) \sim t$ at infinity. Then one can write a  subtracted dispersion relations for $j(t)$:
\eqn\mandelstamargument{
j(t) = \alpha' t + \alpha_0 +{t^2 \over \pi} \int d t' {{\rm Im}[j(t') ] \over (t' - t) t'^2}~.}
From~\pole\ we see that an imaginary part for $j(t)$ at $t_n$ would lead to an unstable particle at $m^2=t_n$. 
Mandelstam concludes that ${\rm Im}[ j(t) ]  = 0$ everywhere and hence the trajectory is linear.
The assumption that $j(t)$ does not grow faster than $t$ can be relaxed while not affecting the conclusion. Indeed, a faster increase at large $t$ would require an extra subtraction and one can then rule out 
a faster growth by demanding the absence of
extra singularities $j(t_n) = n$ in the scattering amplitude which do not have a proper physical interpretation. However, the assumption that $j(t)$ is everywhere analytic is more difficult to justify and contradicts what we expect in Yang-Mills theory (see Fig.~1).
One difficulty is the possibility of level-crossing phenomena: if different trajectories intersect at some complex value of $t$ there can be non-analytic behavior in the individual $j(t)$
yet preserving analyticity of the amplitude.  An example of this phenomenon is given in section 4 of~\KorchemskyRC.
There could also be first-order phase transitions: the asymptotic form of the amplitude (at large $s$) at different values of $t$ could be given by different analytic expressions in different regions of the complex $t$ plane.\foot{The simplest example of this phenomenon is a sum of two Veneziano amplitudes  with different slopes. This leads to an effective $j(t)$ on the real line which is continuous but not differentiable. In fact, such sums can serve as a toy model for Yang-Mills theory so this example is not necessarily esoteric. }
Therefore, the analytic continuation in $t$ of the asymptotic estimate~\definitionRegge\ does not need to coincide with the asymptotic form of the amplitude at different values of $t$.
The analytic structure of $j(t)$ is therefore not constrained.

An additional classic argument regarding $j(t)$ was given in~\FreundHW, where it was claimed that 
$j(t)$ grows faster than $\sqrt t$ as $t\rightarrow\infty$ (this was derived under various additional assumptions). We will see that this is indeed a corollary of the consistency conditions explained above.
(The argument of~\FreundHW\ had to do with the width of ultra-heavy resonances, which we do not discuss here.)

\appendix{B}{Argument For Infinitely Many Particles From Euclidean QFT}

Consider any Large N theory and let $O_{\mu_1...\mu_s}$ be a single-trace operator in the symmetric traceless representation. The theory is assumed to be confining and 
therefore we assume the theory contains stable resonances $\phi^i_{\mu_1...\mu_s}$ with mass squared $(m_s^i)^2$. The equations of motion of such free particles are 
$$\square \phi^i_{\mu_1...\mu_s}=(m_s^i)^2\phi^i_{\mu_1...\mu_s}~,$$
$$\del^{\mu_1}\phi^i_{\mu_1...\mu_s}=0~.$$
The second equation is necessary for positivity of the energy.  $\phi^i_{\mu_1...\mu_s}$ is traceless.

The operator $O_{\mu_1...\mu_s}$  can only overlap with particles of lower spin. This is because to overlap with a particle with higher spin we would need to contract with momentum and hence get a vanishing result. 
$$O_{\mu_1...\mu_s}=\left[p_{\mu_1}\cdots p_{\mu_s}-{\rm traces} \right]\sum_{i_0}C^{i_0}\phi^{i_0}+\cdots +\sum_{i_s}C^{i_s} \phi^{i_s}_{\mu_1,...,\mu_s}~.$$

Because the $\phi$ are free resonances we have 
\eqn\prop{
\langle\phi_{\mu_1 ... \mu_s}(p)\phi_{\nu_1 ... \nu_s }(-p)\rangle={{\rm Pol}(p_\mu)\over p^2-m_s^2}~.
}

It is therefore evident that if there are only finitely many resonances, the correlation function $\langle O_{\mu_1...\mu_s}O_{\mu_1...\mu_s}\rangle$
is a rational function. On the other hand, at infinite momentum, we should approach the CFT correlation function, which contains a branch cut. 
From this one can conclude that there are infinitely many scalar operators.

We can say more than that by consider the energy-moment operator. Note that the energy-momentum operator is conserved but not traceless.
The most general expression for it is 

$$T_{\mu\nu}=(p_{\mu}p_{\nu}-p^2\eta_{\mu\nu})\sum_{i_0}D^{i_0}\phi^{i_0}+\sum_{i_2} D^{i_2}\phi^{i_2}_{\mu\nu}$$
We have not included spin one particles because they would appear as $p_\mu \rho_\nu+p_\nu\rho_\mu$, but conservation would imply $m^2_\rho=0$ and therefore 
only massless vector fields can appear. But then the structure above would not be gauge invariant so we conclude that spin one particles do not contribute.

Let us  analyze the propagator $\langle \phi_{\mu\nu}(p)\phi(-p)\rangle$. By the tracelessness and conservation it has to be trivial. Therefore, in the computation of $\langle T_{\mu\nu}T_{\rho\sigma}\rangle$ we have contributions from the spin 0 propagators, which give rise to the tensor structure 
$$(p_{\mu}p_{\nu}-p^2\eta_{\mu\nu})(p_{\rho}p_{\sigma}-p^2\eta_{\rho\sigma})A(p^2)~,$$
and from the spin2-spin2 propagators, which give rise to the tensor structure (which is uniquely fixed by tracelessness and conservation)
$$  P_{\mu\nu\rho\sigma}(p) B(p^2)~.  $$
Here $$P_{\mu\nu\rho\sigma}(p)=(p_{\mu}p_{\nu}-p^2\eta_{\mu\nu})(p_{\rho}p_{\sigma}-p^2\eta_{\rho\sigma})+{D-1\over 2}\left[(p_{\mu}p_{\rho}-p^2\eta_{\mu\rho})(p_{\nu}p_{\sigma}-p^2\eta_{\nu\sigma})+(\mu\leftrightarrow \nu)\right]$$
One can think of $P_{\mu\nu\rho\sigma}$ as the linearized Weyl tensor.
In the ultraviolet CFT, $B(p^2)\sim \log(p^2)$ (the coefficient being the c-anomaly). Therefore, there must be infinitely many spin-2 particles. In addition, the structure $A(p^2)$ should go to zero in the deep UV. 

Therefore we also proved the existence of infinitely many spin-2 particles. 

Similarly, by matching the sum over resonances to the CFT behavior in the UV one can find an asymptotic behavior of the integrated spectral density. This would be analogous to the ideas of~\PappadopuloJK . 

\appendix{C}{More Solutions to $(3.8)$ }

Here we discuss various additional distributions, not of the type~\uniquesolution, and their physical interpretation. 
An interesting distribution in~\logA\ is $$\rho(t,z,\bar z)=C \log(t)\delta^{(2)}(z-\Lambda_*)~,$$
where $\Lambda_*$ is some scale and we take $\log F=0$. This leads at large $s,t$ to 
\eqn\logtra{\log A = C\log(t)\log(-s)~, }
i.e. a logarithmic trajectory, $A\sim (-s)^{C\log(t)}$. 

This is precisely the behaviour of the Coon amplitude \CoonYW . The amplitude is given by \FairlieAD
\eqn\coon{A(s,t)\sim \prod_{r=0}^\infty {\left((\sigma-1)(s-m^2)+1\right)\left((\sigma-1)(t-m^2)+1\right)-\sigma^r\over \left((\sigma-1)(s-m^2)+1-\sigma^r\right)\left((\sigma-1)(t-m^2)+1-\sigma^r\right) } }
with poles at $p_r=m^2+{\sigma^r-1\over \sigma-1}$. At $\sigma=1$ this amplitude reduces to the Veneziano amplitude. At $\sigma>1$ it is non-unitary and at $\sigma<1$ there is an accumulation point for the poles. Let us proceed with the case $\sigma>1$ even though it is non-unitary. If we set $t=p_k$ then the zeros in $s$ are at $s_r=m^2+{\sigma^{r-k}-1\over \sigma-1}$ for $r=0,...,\infty$. Starting from $r=k,...,\infty$ these zeros exactly cancel the poles in $s$. Hence, the excess zeros are given by 
\eqn\excesscoon{s_r=m^2+{\sigma^{r-k}-1\over \sigma-1}~,\qquad r=0,..,k-1~.}
Since $\sigma>1$, for large $k$ they accumulate at $m^2-{1\over \sigma-1}$. At large $k$ the number of excess zeros is ${1\over \log(\sigma)}\log t$ and therefore the Regge behaviour is as in~\logtra\ with $C={1\over \log(\sigma)}>0$.

Another interesting instance of the Coon amplitude is when $\sigma<1$. The poles have an accumulation point and thus the amplitude is not meromorphic but it is formally unitary. The discussion in this case is more subtle. If we take $t=p_k$ then the excess zeros in $s$ are at~\excesscoon\ and hence they go to $-\infty$ for large $k$. The number of excess zeros, $k$, can be written as ${1\over \log(\sigma)} \log\left(1-(1-\sigma)(t-m^2)\right)$ and the Regge behaviour is thus 
\eqn\unicoon{\log A={1\over \log(\sigma)} \log\left(1-(1-\sigma)(t-m^2)\right)\log(-s)~.}

When $t$ crosses the branch point $t=m^2+{1\over 1-\sigma}$ there is no longer a notion of paired and free zeros in $s$. Indeed all the zeros are to the right of the accumulation point while the poles are to the left. This contradicts the fact that between any two poles there must be a zero in unitary amplitudes. This suggests that the product~\coon\ is badly behaved past the accumulation point. 
In addition, if we consider~\unicoon\ and attempt to continue $t$ past the branch point $m^2+{1\over 1-\sigma}$ we get $\log A \sim \log(-s)\log(-t)$ for large  positive $s,t$. This has a double cut, which is not allowed in meromorphic amplitudes, but this is consistent with amplitudes that have an accumulation point. 
Furthermore, the coefficient of $\log(-s)\log(-t)$ is ${1\over \log\sigma}$ which is negative in this case and again inconsistent with unitarity although the amplitude with $\sigma<1$ is nominally unitary, namely all the residues are polynomials which are given by a sum of Legendre polynomials with positive coefficients \FairlieAD . 
Because of all of these sicknesses of the $\sigma<1$ case, we do not discuss further the unitary Coon amplitude and its variations in this paper. Taking $\sigma \to 0$ limit leads to the toy model amplitude \toysolution\ considered in the introduction.

\appendix{D}{Review of the Veneziano Amplitude}

Let us quickly review the most famous solution to the problem outlined in the introduction, namely the Veneziano amplitude \VenezianoYB. We consider scattering of four particles of mass $m^2$ and assume that there are no $u$-channel poles. The Veneziano amplitude takes the following form
\eqn\Veneziano{
A(s,t) =  {\Gamma(- \alpha_0 - s) \Gamma(- \alpha_0 - t) \over \Gamma( - 2 \alpha_0 - s - t)} .
}

This amplitude is manifestly crossing symmetric and has the form \definitionRegge\ with $j(t) =t + \alpha_0 $.
The poles are at $s,t =n-\alpha_0$ with $n=0,1,2,\dots$ in accordance with the masses of particles that are being exchanged.

Unitarity implies that the residues at the location of particles can be decomposed in terms of Legendre polynomials with positive coefficients
\eqn\unitarityV{\eqalign{
-{\rm Res}_{s = \alpha_0 + n} [A(s , t)] &= \sum_{j=0}^{n} c_{n,j} P_j\left(1 + {2 t \over \alpha_0 + n - 4 m^2}\right)\ ,\qquad c_{n,j} \geq 0 \ , \cr{\rm where}\quad
P_j (x ) &= \,_2 F_1 (- j , j + D - 3 , {D-2 \over 2},{ 1 - x \over 2 })~.
}}

Imposing the positivity of the residues constrains the mass of the external particle, $m^2$, the Regge intercept $\alpha_0$, and dimensionality of spacetime $D$. It is very easy to work out the constraints from the first few levels as we discuss briefly below. When $\alpha_0 = - m^2 = 1$ one reproduces the famous bound $D \leq 26$. Another convenient choice is $\alpha_0 = m^2 = 0$, in this case we get $D \leq 10$. These are of course the critical dimensions of string theory.

Notice that positivity of $c_{n,j}$ together with positivity of the Legendre polynomials for $t>0$ implies that the amplitude becomes large for $s,t \gg 1$. The delicate cancelations between sign-alternating Legendre polynomials that are possible for $t<0$ cannot occur for positive $t$.

To study the asymptotic properties of the Veneziano amplitude it is useful to rewrite it as follows
\eqn\rewrite{\eqalign{
A(s,t) &= - {\pi \sin(\pi(2 \alpha_0 + s + t)) \over \sin(\pi(\alpha_0 + s)) \sin(\pi ( \alpha_0 + t))} A^{\rm excess}(s,t)  , \cr
A^{\rm excess}(s,t) &= {\Gamma(s + t + 2 \alpha_0 + 1) \over  \Gamma(s +  \alpha_0 + 1) \Gamma(t + \alpha_0 + 1) } \ .
}}

By construction, $A^{\rm excess}(s,t)$ is equal to the residue of the amplitude whenever $s$ or $t$ take the values $\alpha_0 + n$. At these points it becomes a polynomial which has a decomposition in terms of Legendre polynomials with positive coefficients when the amplitude is unitary. Notice also that it is crossing-symmetric so that for $s,t =\alpha_0 + {\rm integer}$ we have a polynomial crossing equation
\eqn\polycross{
{\rm Res} [ A(n+ \alpha_0, m+\alpha_0)] =  {\rm Res} [ A(m+ \alpha_0, n+\alpha_0)] \ .
}
.

For generic $s,t$, the function $A^{\rm excess}(s,t)$ is related to the discontinuity of the amplitude. Indeed, notice that upon taking $s \to s(1 \pm i \eps)$ with large positive $s$ we have
\eqn\simplething{
\lim_{|s|\to\infty}\left.- {\pi \sin(\pi(2 \alpha_0 + s + t)) \over \sin(\pi(\alpha_0 + s)) \sin(\pi ( \alpha_0 + t))}\right|_{s=|s|(1 \pm i \eps)}\  =\  {\pm 2 \pi i \over 1 - e^{\pm 2 \pi i (t + \alpha_0)}} \ .
}

From this we see that for $s,t \gg 1$
\eqn\definition{
A(s(1+i \eps), t(1+ i \eps)) = {\rm Disc}_s A(s,t) = {\rm Disc}_t A(s,t) =  2 \pi i \  A^{\rm excess}(s,t) \ .
}
where the discontinuity is defined as usual
\eqn\discdef{
{\rm Disc}_s A(s,t)  = A(s(1+i \eps), t(1+ i \eps)) - A(s(1-i \eps), t(1+ i \eps))
}
and analogously in the $t$-channel.

In the language of the body of the paper, $A^{\rm excess}(s,t)$ comes from the distribution of the excess zeros. We focus on the leading piece in the limit $s,t \gg 1$ which using \rewrite\ is given by \gm. Namely, 
\eqn\limitform{
\log A^{\rm excess} (s,t) = (s+t) \log (s+t) - s \log s - t \log t .
}

This formula is, of course, reminiscent of the fixed angle scattering analysis by Gross and Mende~\GrossKZA. In our case the amplitude is exponentially large instead of being exponentially suppressed because we are in a different kinematical region.

\listrefs

\bye